\documentclass[a4paper,11pt]{article}

\usepackage[dvipsnames]{xcolor}
\usepackage{tikz}
\usepackage{jheppub}
\usepackage{graphicx}
\usepackage{amsmath,amssymb,amsthm,amsfonts}
\usepackage{slashed}
\usepackage{braket}
\usepackage[export]{adjustbox}
\usepackage{enumitem}
\usepackage{placeins}
\usepackage[normalem]{ulem}
\usepackage{multirow}
\usepackage{mathrsfs}
\usepackage{amsthm}
\usepackage{dsfont}
\usepackage{xcolor}
\usepackage{tcolorbox}
\usepackage{verbatim}
\usepackage[T1]{fontenc}
\usepackage[utf8]{inputenc}
\usepackage{lmodern}
\usepackage{array}
\usepackage{wrapfig}
\usepackage{tabu}
\usepackage{longtable}
\usepackage{float}
\usepackage{caption}
\usepackage{subcaption}
\usepackage{tikz,tikz-3dplot}
\usepackage[compat=1.1.0]{tikz-feynman}
\usetikzlibrary{calc,arrows.meta,shapes.geometric}
\usepackage{soul}
\usepackage[usestackEOL]{stackengine}
\usepackage{cancel}
\usepackage{hyperref}
\usepackage{cleveref}
\allowdisplaybreaks

\definecolor{niceGreen}{rgb}{0.015,0.386, 0.0273}
\definecolor{deepBlue}{RGB}{0, 20, 158}
\definecolor{deepRed}{RGB}{158, 0, 0}

\hypersetup{
    colorlinks=true,
    linkcolor=niceGreen,
    citecolor=niceGreen,
    filecolor=niceGreen,      
    urlcolor=niceGreen
    }

\def\sdot{\!\cdot\!}

\newcommand{\Gsubgammagraph}{
\begin{minipage}{2.25cm}
\begin{tikzpicture}
    \draw[fill=black] (0,-2) circle (0.1) ;
    \node at (0.65,0.15) {$\hat{\gamma}$};
    \node at (0.45,-2.25) {$v_{\gamma}$};
    \draw[ thick, -] (0.26,-0.15) to[out=-50,in=50] (0.09,-1.96);
    \draw[ thick, -] (-0.26,-0.15) to[out=-130,in=130] (-0.09,-1.96);
    \node at (0.05,-1.2) {\LARGE$\cdots$};
    \node at (-0.824,-1.2) {\footnotesize{$k_1$}};
    \node at (0.87,-1.2) {\footnotesize$k_m$};
    \draw[ thick, -,deepRed][fill=white,postaction={pattern=north east lines, pattern color=deepRed}] (0,0) circle (0.30) ;
\end{tikzpicture}
\end{minipage}
}

\newcommand{\subgammagraph}{
\begin{minipage}{2.25cm}
\begin{tikzpicture}
    \foreach \angle in {-150,-125,-100,-50} {
       \draw[thick]
         ($(0,0) + (\angle:0.6)$) -- ++(\angle:0.6);
     }
    \foreach \angle in {-90,-75,-60} {
       \fill ($(0,0) + (\angle:0.78)$) circle (0.03);
     }
    \draw[ thick, -,deepBlue][fill=white,postaction={pattern=north east lines, pattern color=deepBlue}] (0,0) circle (0.60) ;
    \node at (-1.2,-0.8) {\footnotesize{$p_1$}};
    \node at (-0.8,-1.2) {\footnotesize{$p_2$}};
    \node at (-0.2,-1.4) {\footnotesize{$p_3$}};
    \node at (0.9,-1.1) {\footnotesize$p_n$};
\end{tikzpicture}
\end{minipage}}
    
\newcommand{\GGraph}{
\begin{minipage}{2.5cm}
\begin{tikzpicture}
    \draw[thick, -] (0.424,0.424) to[out=60,in=-40] (0.106,1.894);
    \draw[thick, -] (-0.424,0.424) to[out=120,in=-140] (-0.106,1.894);
    \draw[ thick, -,deepRed][fill=white,postaction={pattern=north east lines,
        pattern color=deepRed}] (0,2) circle (0.30);
    \node at (0.05,1.1) {\LARGE$\cdots$};
    \node at (-0.824,1.1) {\footnotesize{$k_1$}};
    \node at (0.87,1.1) {\footnotesize$k_m$};
    \foreach \angle in {-150,-125,-100,-50} {
       \draw[thick]
         ($(0,0) + (\angle:0.6)$) -- ++(\angle:0.6);
     }
     \foreach \angle in {-90,-75,-60} {
       \fill ($(0,0) + (\angle:0.78)$) circle (0.03);
     }
     \draw[ thick, -,deepBlue][fill=white,postaction={pattern=north east lines,
        pattern color=deepBlue}] (0,0) circle (0.60);
    \node at (0.95,0.15) {};
    \node at (-1.2,-0.8) {\footnotesize{$p_1$}};
    \node at (-0.8,-1.2) {\footnotesize{$p_2$}};
    \node at (-0.2,-1.4) {\footnotesize{$p_3$}};
    \node at (0.9,-1.1) {\footnotesize$p_n$};
    \node at (0.65,2.15) {$\hat{\gamma}$};
\end{tikzpicture}
\end{minipage}}

\newcommand{\BubbleGraph}{
\begin{minipage}{2.5cm}
\begin{tikzpicture}
\draw[ thick, -] (0.5,0) arc[start angle=0, end angle=-180,radius=0.5];
\draw[ thick, -] (-0.5,0) -- (-1.1,0);
\draw[ thick, -] (0.5,0) -- (1.1,0);
\draw (0.8,0.2) node{${\scriptstyle {p}}$};
\draw (-0.8,0.2) node{${\scriptstyle {p}}$};
\fill[ thick, -] (0.5,0) circle (0.05);
\fill[ thick, -] (-0.5,0) circle (0.05);

\draw[ thick, -] (0,0.5) circle (0.2);

\draw[ thick, -] (-0.5,0) arc[start angle=180, end angle=112, radius=0.5];
\fill[ thick, -] (0.2,0.45) circle (0.035);
\fill[ thick, -] (-0.2,0.45) circle (0.035);
\draw[ thick, -] (0.5,0) arc[start angle=0, end angle=68, radius=0.5];
\end{tikzpicture}
\end{minipage}
}

\newcommand{\GsubGEx}{
\begin{minipage}{1cm}
\begin{tikzpicture}
\draw[ thick, -] (0,0) circle (0.25) ;
\draw[fill=black] (0,-1.5) circle (0.05) ;
\node at (0.45,0) {$\hat{\gamma}$};
\node at (0.4,-1.5) {$v_{\gamma}$};
\draw[ thick, -] (0.18,-0.18) to[out=-60,in=40] (0.04,-1.48);
\draw[ thick, -] (-0.18,-0.18) to[out=-120,in=140] (-0.04,-1.48);
\end{tikzpicture}
\end{minipage}
}

\newcommand{\GamEx}{
\begin{minipage}{2cm}
\begin{tikzpicture}
\draw[ thick, -] (0.4,0) arc[start angle=0, end angle=-180,radius=0.5];
\draw[ thick, -] (-0.6,0) -- (-1.2,0);
\draw[ thick, -] (0.4,0) -- (1,0);
\draw (0.8,0.2) node{${\scriptstyle {p}}$};
\draw (-0.9,0.2) node{${\scriptstyle {p}}$};
\fill[ thick, -] (0.4,0) circle (0.05);
\fill[ thick, -] (-0.6,0) circle (0.05);
\end{tikzpicture}
\end{minipage}
}

\tikzstyle{pink} = [dashed, color=RubineRed]

\preprint{\phantom{????}}

\title{On divergences in a four-derivative scalar field theory}

\author[a,b]{Maegan Anderson,}
\emailAdd{m.m.h.anderson@ed.ac.uk}

\author[b]{Sam Bateman,}
\emailAdd{sam.bateman@ed.ac.uk}

\author[b]{Franz Herzog,}
\emailAdd{fherzog@ed.ac.uk}

\author[b,c]{Neil Turok.}
\emailAdd{neil.turok@ed.ac.uk}

\affiliation[a]{School of Mathematics and Maxwell Institute for Mathematical Sciences,\\The University of Edinburgh, Edinburgh EH9 3FD, Scotland, UK}

\affiliation[b]{Higgs Centre for Theoretical Physics, School of Physics and Astronomy,\\The University of Edinburgh, Edinburgh EH9 3FD, Scotland, UK}

\affiliation[c]{Perimeter Institute for Theoretical Physics, Waterloo, Ontario N2L 2Y5, Canada
}

\abstract{We perform a detailed diagrammatic analysis of the renormalisation of a family of asymptotically free, shift-symmetric four-derivative scalar field theories introduced by Holdom~\cite{Holdom:2023usn,Holdom:2024cfq}. We extend the renormalisation of the theory from one to three loops using both an $R^*$ method and an asymptotic expansion in momenta. We prove that the Euclidean correlators (or off-shell amplitudes) are IR finite, to all orders in perturbation theory, and derive a non-renormalisation theorem describing the all-order structure of the renormalisation constants. 
In particular, a purely cubic interaction is RG invariant and a perfect square Lagrangian density is preserved under renormalisation. The latter result is due to a Ward identity in a related {\it gravitational} theory -- the conformally flat limit of quadratic gravity (CFQG) (see our companion paper Ref.~\cite{Anderson:2026Rsqaured}). We show that the beta function for the perfect square theory maps exactly to that of an $O(2)$-symmetric, two-derivative, massless $\phi^4$ theory at negative coupling. We verify this relationship explicitly up to three loops and thus determine the beta function and anomalous dimension for both the perfect square theory and CFQG to six loops.}

\begin{document}

\maketitle

\section{Introduction}

Among the set of all quantum field theories (QFTs), asymptotically free theories occupy a special position. They are believed to be UV complete and therefore to possess a continuum limit as their high energy behaviour is weakly coupled and well described by perturbation theory. However, in conventional approaches to QFT, the only known examples of asymptotic freedom in four dimensions are non-abelian gauge theories, including their matter-coupled versions~\cite{Coleman:1973freedom}. Expanding this list of asymptotically free quantum field theories is an important objective which, however, calls for generalising QFTs beyond the current paradigm.

The Ostrogradsky no-go theorem~\cite{Ostrogradsky:1850fid} places one of the strongest known constraints on the space of viable QFTs,  requiring that the Lagrangian depends only on first time derivatives (see, \textit{e.g}, Ref.~\cite{Woodard:2015zca} for a recent review). Weakening this constraint could expand the list of asymptotically free QFTs substantially, to include quadratic gravity for example~\cite{Stelle:1976gc,Fradkin:1981iu,Avramidi:1985ki}, as well as four-derivative scalar field theories~\cite{Holdom:2023usn,Holdom:2024cfq}. Such higher derivative theories might well play a role either in describing quantum gravity or in resolving some of the Standard Model's deep puzzles such as the gauge-gravity hierarchy and the Landau poles in some couplings. Furthermore, such four-derivative field theories are intimately related to Weyl anomalies  arising in the effective action~\cite{Riegert:1984kt,Antoniadis:1991fa} and possible cancellation mechanisms~\cite{Boyle:2021jaz}. Moreover, their inclusion might explain the primordial cosmological perturbations~\cite{Turok:2023amx}, and create UV fixed points in classical gravitational beta functions \cite{Boyle:2025bxf}. These developments suggest that four-derivative theories might play an integral role in extensions of the SM which attempt to consistently incorporate gravity.

Many previous works have explored higher derivative theories, both in quantum mechanics and in quantum field theory. Fradkin and Tseytlin pioneered the calculation of beta functions in four-derivative theories of gravity, including conformal supergravity~\cite{Fradkin:1981hx,Fradkin:1981iu,Fradkin:1985am,Tseytlin:2017qfd}. Subsequently, Tseytlin and co-authors discussed perturbative unitarity for four-derivative scalar theories when interactions are included, pointing out the key problems of correctly identifying the physical observables and asymptotic states~\cite{Adamo:2018srx,Nakach:2018jyu,Tseytlin:2022flu}. 

Ostrogradsky's theorem when combined with the correspondence principle implies that the {\it expectation value} of the Hamiltonian can be arbitrarily negative. In perturbative QFT, however, a key assumption is that all particle states have positive energies, {\it i.e.}, that the {\it eigenvalues} of the Hamiltonian are all positive. Reconciling the two statements requires that negative norm (or ghost) states be included~\cite{Bateman:2026letter}. The danger which then arises is the possibility of obtaining negative transition probabilities.

Holdom studied a general family of renormalisable, shift-invariant four-derivative scalar theories~\cite{Holdom:2023usn,Holdom:2024cfq}.  He found the surprising result that the tree-level elastic cross section is positive in a certain region of coupling constant space. Recently, two of us studied the special theory on the boundary of this region, the ``perfect square'' (PS) theory. For this theory, we were able to extend this positivity result to a very general set of physical observables due to a hidden ghost parity symmetry~\cite{Bateman:2026letter}. In particular, we considered the transition probabilities for arbitrary covariant scattering processes, which includes decay widths, cross-sections, as well as their multi-particle generalisations. These results indicate that the existence of ghosts does not necessarily imply the prediction of negative probabilities.

In a companion paper to this one, we show that PS theory describes the conformally flat limit of quadratic gravity (CFQG), suggesting it is a physically consistent, UV complete theory of quantum gravity, albeit in the limit where gravitons decouple~\cite{Anderson:2026Rsqaured}. 

In this work we investigate the higher-order, perturbative structure of Holdom's family of four-derivative theories in a diagrammatic analysis. Our key results are as follows:
\begin{itemize}
\item We prove a general structural theorem governing the coupling dependence of the renormalisation constants. It shows, in particular that the $n$-point correlation functions are superficially finite for $n$ > 4, as required by renormalisability.
    \item We prove that all correlation functions in these theories are infrared finite in momentum space, off-shell, as a consequence of the momentum dependence of the interaction vertices imposed by shift symmetry. 
    \item For two special cases of Holdom's family of theories, we prove and explicitly check two non-renormalisation results:
    \begin{itemize}
        \item When the quartic coupling is zero, the theory is Galileon invariant. We show that the cubic interaction is renormalisation group (RG) invariant to all orders.  This extends an observation made by Safari {\it et al.} at one loop \cite{Safari:2021ocb}. 
        \item When the Lagrangian density is a perfect square, its form is preserved under RG flow. Holdom observed this behaviour at one loop. In the companion paper~\cite{Anderson:2026Rsqaured}, we explain it as a consequence of the Ward identity associated with diffeomorphism invariance in CFQG, which holds to all orders. Here, we verify the special RG trajectory explicitly up to three loops. 
    \end{itemize}
    \item We show that the PS theory may be formally mapped into a standard two-derivative, $O(2)$-invariant, $\phi^4$ scalar theory at negative coupling, a theory long known to be asymptotically free \cite{Symanzik:1973hx}. We explicitly verify that the two theories have identical beta functions up to three loops. Therefore, we obtain a correspondence between standard $\phi^4$ theory and CFQG which, using known results in $\phi^4$, allows us to predict the beta function of CFQG up to six loops. 
\end{itemize}
To compute the renormalisation constants of the general shift-invariant model we employ the $R^*$ method \cite{Chetyrkin:1982nn,Vladimirov:1979zm,
Chetyrkin:1984xa,Smirnov:1985yck,Larin:2002sc,Kleinert:2001hn,Chetyrkin:2017ppe,Herzog:2017bjx,deVries:2019nsu,Beekveldt:2020kzk,Henriksson:2025hwi} . This is an extension of Bogloliubov-Parasiuk-Zimmermann-Hepp renormalisation where Bogoliubov's $R$ operation is promoted to an $R^*$ operation, which beyond UV also handles IR divergences in off-shell quantities. The real power of $R^*$ however lies in the fact that it simplifies the computation of UV divergences. This is achieved by re-routing or nullifying external momenta of a Feynman diagram, such that the integral is easier to compute. While this procedure does not change the overall UV divergence in the minimal subtraction (MS) of dimensional regularisation, it does introduce unphysical IR divergences, which $R^*$ is equipped to subtract. 

While the $R^*$ method has been successfully applied to high-loop calculations, \emph{e.g.}, 5-loop renormalisation of leading twist operators in QCD \cite{Falcioni:2023luc,Falcioni:2023vqq,Falcioni:2024qpd,Falcioni:2024xav,Falcioni:2024xyt}, or high- dimensional conformal operators in a general scalar EFT  \cite{Henriksson:2025hwi}, the application of $R^*$ to the present theory turns out to be intricate. This is due to both the large positive powers of momenta in the vertices and the large negative powers in the propagators. The former require repeated application of Taylor operators to lower the degree of divergence of subgraphs. The latter lead to high degree spurious IR divergences which, in turn, lead to long recurrences. Computing the Feynman integrals in the theory is also challenging, with many terms per diagram and many integrations by parts required. Despite these theories involving just one scalar field, calculations in them are every bit as challenging as those in Yang Mills theory, albeit without the intricacies of vector fields and spinor and colour algebra. 

Besides their possible physical relevance, these four derivative theories serve as a useful laboratory for developing new calculational methods for higher derivative operators, such as are needed in QCD for the operator product expansion and in effective field theories more generally. The four-derivative theories we study here rival Yang-Mills theory in terms of both computational complexity as well as mathematical elegance in the sense of symmetries, Ward identities and non-trivial physical properties. 

The paper is structured as follows. The general four-derivative Lagrangian, its corresponding Feynman rules and our renormalisation conventions are described in  \cref{sec:Background}. The structural theorem and our non-renormalisation results are explained in \cref{sec:Non-renormalisation theorem}. Our proof of the IR finiteness of off-shell correlators is presented in \cref{sec:IRfiniteness}, along with an accompanying example. The special case of the perfect square theory, its embedding within an $O(1,1)$-symmetric two-derivative theory and their beta function equivalence with standard $\phi^4$ theory at negative coupling are explained in \cref{sec:Oclassicaly}. Explicit computations up to three loops and six-loop predictions for the beta functions of PS theory and CFQG, based on known results for $\phi^4$, are presented in \cref{sec:ComputationsandResults}. Section \ref{sec:Conclusion} concludes. 

\section{The four-derivative theory}
\label{sec:Background}

We study a renormalisable interacting theory consisting of four-derivative scalars which have classical mass dimension zero in four dimensions. The free field theory is described by the action
\begin{equation}\label{eq:FreeFieldDimZeroLagrangian}
    \mathcal{S} = - \frac12\int d^4x \; (\square \sigma)^2 \,,
\end{equation}
giving rise to the fourth order field equation $\square^2\sigma=0$. This theory is known as the massless dipole ghost. Such field equations have a long history dating back to Bhabha~\cite{bhabha1950nuclear} and Heisenberg~\cite{heisenberg1957lee}, and have recently garnered attention as toy models of quadratic gravity~\cite{Bender:2007wu,Bender:2008gh,Donoghue:2021cza,Donoghue:2021eto,Donoghue:2021meq,Holdom:2015kbf,Holdom:2021hlo,Holdom:2023usn,Holdom:2024cfq}. The free theory yields the propagator
\begin{equation}
    \begin{tikzpicture}[scale=0.8,baseline={([yshift=-1.77ex]current bounding box.center)}]
        \draw[thick] (-1,0)--(1,0);
        \node at (0,0) [above] {\scriptsize$p$};
    \end{tikzpicture} \, = -\frac{i}{(p^2+i\epsilon)^2} \,,
\end{equation}
where the $i\epsilon$ prescription is fixed by the spectral condition or, equivalently, by analytic continuation from the Euclidean propagator. As one can easily check, the Euclidean action corresponding to (\ref{eq:FreeFieldDimZeroLagrangian}) is positive. 

\subsection{Interacting theory and Feynman rules}

The free action \eqref{eq:FreeFieldDimZeroLagrangian} is invariant under the constant shift $\sigma \mapsto \sigma + c$. It was noted in ref.~\cite{Tseytlin:2022flu} that this shift symmetry tames the IR behaviour of the theory. The marginal, shift invariant interactions were categorised in \cite{Safari:2021ocb}. In four dimensions, there are two such operators, which are respectively cubic and quartic in $\sigma$. The most general shift invariant interacting theory is then described by the Lagrangian
\begin{equation}\label{eq:LagArbCoupl}
    \mathcal{L} = - \frac12 \sigma\Box^2 \sigma + \lambda_3 (\partial_\mu \sigma \partial^\mu \sigma)\Box \sigma + \lambda_4 (\partial_\mu \sigma \partial^\mu \sigma)^2 \,,
\end{equation}
where $\lambda_3$ and $\lambda_4$ are dimensionless couplings. The corresponding quantum field theory was first investigated in detail by Holdom in \cite{Holdom:2023usn,Holdom:2024cfq}. For on-shell quantities such as the cross section, he found it convenient to include the shift-invariant mass term $m^2(\partial\sigma)^2$, sending the mass to zero at the end of calculations. For the off-shell quantities we shall calculate here, we find it unnecessary to include a mass regulator, and for simplicity we shall focus on the purely massless theory. The Feynman rules for the cubic and quartic vertices are 
\begin{align}
    \begin{tikzpicture}[baseline={([yshift=-0.5ex]current bounding box.center)}]
    \draw[thick] (0,0)--(-0.5,0) node[left] {${\scriptstyle {p_1}}$};
    \draw[thick] (0,0)--(60:0.5) node[right] {${\scriptstyle {p_2}}$};
    \draw[thick] (0,0)--(-60:0.5) node[right] {${\scriptstyle {p_3}}$};
    \filldraw (0,0) circle (2pt);
    \end{tikzpicture}
    = 2i\lambda_3 V_3(p_1,p_2,p_3)\, ,
    \qquad
    \begin{tikzpicture}[baseline={([yshift=-0.5ex]current bounding box.center)}]
    \draw[thick] (0,0)--(-135:0.5) node[left]{${\scriptstyle {p_2}}$};
    \draw[thick] (0,0)--(135:0.5) node[left]{${\scriptstyle {p_1}}$};
    \draw[thick] (0,0)--(45:0.5) node[right]{${\scriptstyle {p_3}}$};
    \draw[thick] (0,0)--(-45:0.5) node[right]{${\scriptstyle {p_4}}$};
    \filldraw (0,0) circle (2pt);
    \end{tikzpicture}
    = 8i\lambda_4 V_4(p_1,p_2,p_3,p_4) \,,
\end{align}
where
\begin{align}
    V_3(p_1,p_2,p_3) &= p_1\sdot p_2\, p_3^2 + p_1\sdot p_3\, p_2^2 + p_2\sdot p_3\, p_1^2\,,
    \\
    V_4(p_1,p_2,p_3,p_4) &= (p_1 \sdot p_2 )\,(p_3 \sdot p_4) + (p_1 \sdot p_3) \, (p_2 \sdot p_4) + (p_1 \sdot p_4 ) \, ( p_2 \sdot p_3) \,,\label{eq:V4}
\end{align}
define the momentum-dependent vertex factors. Note that at every vertex we have momentum conservation, so we can also rewrite the cubic vertex factor as 
\begin{equation}\label{eq:V3_Gram}
    V_3(p_1,p_2,-p_1-p_2) = 2(p_1 \sdot p_2)^2 - 2p_1^2p_2^2 = -2\det\begin{pmatrix} p_1^2 & p_1\sdot p_2 \\ p_2\sdot p_1 & p_2^2 \end{pmatrix} \,,
\end{equation}
which has the structure of a Gram determinant for the two independent momenta. This structure implies that the cubic vertex scales quadratically with each of the momenta. However, a similar property does not hold for $V_4$ which scales linearly with each momentum.

\subsection{Renormalisation}\label{sec:Rernormalisation}

To renormalise this theory, we use dimensional regularisation and the minimal subtraction scheme. Taking \eqref{eq:LagArbCoupl} to be the bare Lagrangian, we introduce the renormalised field and couplings 
\begin{equation}\label{eq:FieldBare}
    \sigma^{b} = Z_\sigma^{\frac{1}{2}} \sigma^{r} \,,
    \qquad
    \lambda_{3}^{b} =(4\pi)\mu^{\frac{\varepsilon}{2}}Z_{3}  \lambda_{3}^{r} \,,
    \qquad
    \lambda_{4}^{b} = (4\pi)^2 \mu^{\varepsilon} Z_{4} \lambda_{4}^{r} \,,
\end{equation}
where the 't Hooft scale $\mu$ accounts for the missing mass dependence in dimensional regularisation in 
\begin{equation*}
    d = 4 - \varepsilon \,.
\end{equation*}
We have introduced factors of $(4\pi)$ to simplify our results in later sections, as such factors are generated at each loop order. Writing the bare Lagrangian in terms of renormalised quantities then yields
\begin{equation}\label{eq:RenromalisedInteractingLagrangian}
    \mathcal{L} = -\frac{1}{2} Z_\sigma \sigma \Box^2 \sigma +  \lambda_3 (4\pi)\mu^{\frac{\varepsilon}{2}} Z_1   (\partial_\mu \sigma   \partial^\mu \sigma) \Box \sigma +   \lambda_4 (4\pi)^2 \mu^{\varepsilon}Z_2 (\partial_\mu \sigma  \partial^\mu \sigma)^2 \;,
\end{equation}
where we have also introduced $Z_1 = Z_\sigma^{3/2}Z_{3}$ and $ Z_2 =Z_\sigma^2 Z_{4}$ and we have dropped the superscript $r$ for convenience. The scale invariance of the theory is broken by loop corrections. The scale dependence of the couplings is characterised by the beta functions
\begin{equation}
     \beta_i = \mu \frac{d}{d\mu} \lambda_i =  \sum_{\ell = 0}^{\infty}\beta_i^{(\ell)}(\lambda_3,\lambda_4) \,,
\end{equation}
where $\beta^{(\ell)}$ is the $\ell$-loop beta function coefficient. It follows from the $\mu$-independence of the bare couplings that
\begin{equation}\label{eq:fullbetai}
    \begin{pmatrix}
        \beta_3 \\ \beta_4
    \end{pmatrix}
    = -\varepsilon
    \begin{pmatrix}
        1+A_{33} & A_{34}\\ A_{43} & 1+A_{44}
    \end{pmatrix}^{-1}
    \begin{pmatrix}
        \frac{1}{2}\lambda_3 \\ \lambda_4
    \end{pmatrix},
\end{equation}
where $A_{ij}=\lambda_i\,\partial_{\lambda_j}\log Z_i$, for $i=3,4$, and the renormalisation factors $Z_i = Z_i(\lambda_3,\lambda_4,\varepsilon)$ are power series in $1/\varepsilon$ whose coefficients involve powers of the renormalised couplings. 

An $\ell$-loop Feynman diagram with $n$ external legs and $n_i$ powers of $\lambda_i$ scales like $\lambda_3^{n_3}\lambda_4^{n_4}$, with $n_3$ and $n_4$ non-negative integers. For any such diagram, we have the identity $3n_3+4n_4=n+2E$ where $E$ is the number of internal edges. Euler's identity then yields
\begin{equation}\label{eq:numberoflegsforloops}
    n_3 + 2n_4 = 2(\ell-1) + n\,.
\end{equation}
Using this relation we can express the loop expansion for the $n$-point amputated renormalised (by which we mean expressed in terms of renormalised coupling, but without counterterms) 1PI correlator in momentum space (henceforth also referred to as the $n$-vertex, or the self-energy for $n=2$) as
\begin{equation}\label{eq:Gamma_loop_expansion}
    \Gamma_n^r (\lambda_3,\lambda_4) = \sum_{\ell=0}^{\infty} \sum_{\substack{n_3 + 2n_4 \\ = 2(\ell-1) +n}} \left(\lambda_3\right)^{n_3}\left(\lambda_4\right)^{n_4}\,\Gamma_n^{(n_3,n_4)} \,.
\end{equation}
Here $\Gamma_n^{(n_3,n_4)}$ is used to indicate the corresponding coefficient for given powers of $n_3$ and $n_4$ vertices. Note that at leading order $\Gamma_n^{(n_3,n_4)}$ is just the Feynman rule of the vertex, this is apparent by writing   
\begin{equation}
    \Gamma_i=\lambda_iV_i+ \delta\Gamma_i\,,
\end{equation}
where $\delta\Gamma_i$ collects the loop corrections and $V_i$ is the vertex Feynman rule. Including renormalisation counterterms we obtain the finite or fully renormalised $n$-vertex 
\begin{equation}\label{eq:bare_correlator}
\Gamma_n^R (\lambda_3,\lambda_4) = Z_\sigma^{n/2} \Gamma_n^r (Z_3\lambda_3,Z_4\lambda_4)\,.
\end{equation}
From the finiteness one can then extract the following expressions for the two renormalisation factors in the minimal subtraction scheme, namely 
\begin{align}
\label{eq:deltaZ1}
    Z_1  - 1&=-\frac{1}{\lambda_3 V_3} K_\varepsilon \left( Z^{3/2}_\sigma\, \delta\Gamma_3^r(Z_3\lambda_3,Z_4\lambda_4)\right)= \sum_{\ell=1}^{\infty} \sum_{\substack{n_3 + 2n_4 \\ = 2\ell+1}} \left(\lambda_3\right)^{n_3-1}\left(\lambda_4\right)^{n_4}\,Z_1^{(n_3,n_4)}\,,\\
    \label{eq:deltaZ2}
    Z_2 -1  &=-\frac{1}{\lambda_4 V_4} K_\varepsilon \left( Z^{2}_\sigma \, \delta\Gamma_4^r(Z_3\lambda_3,Z_4\lambda_4)\right)= \sum_{\ell=1}^{\infty} \sum_{\substack{n_3 + 2n_4 \\ = 2\ell+2}} \left(\lambda_3\right)^{n_3}\left(\lambda_4\right)^{n_4-1}\,Z_2^{(n_3,n_4)}\,,
\end{align}
where $V_i$ removes kinematic dependence and $K_\varepsilon$ is the operator which only picks up poles in $\varepsilon$, essentially removing finite contributions. Moreover, $Z_i^{(n_3,n_4)}$ indicates the corresponding coefficient in the renormalisation factor for a given graph with a given number, $n_3$ and $n_4$, of vertices. We expand the counterterms in the number of loops $\ell$ as 
\begin{eqnarray}
    Z_i - 1 =   \delta_i \,, \qquad  \text{where} \qquad  \delta_i = \sum_{\ell=1}^\infty \delta_i^{(\ell)} \,,
\end{eqnarray}
and the counterterms $\delta_i^{(\ell)}$ are power series in positive powers of $1/\varepsilon$, whose coefficients we find to involve only positive powers of the renormalised couplings.  We remark that this positivity is a non-trivial property and it guarantees that both limits $\lambda_3\to 0,\lambda_4\to 0$ are well-defined leading to independently renormalisable QFTs.

Using the full expression for the beta functions, given by \cref{eq:fullbetai}, ensures that all poles in $\varepsilon$ cancel appropriately from the counterterms. This calculation provides a powerful check on our results. The contribution to each beta function at any given loop order arises from the coefficient of $1/\varepsilon$ in the corresponding counterterm. This can be seen as follows:
\begin{equation}\label{eq:beta_coeffs}
     \beta^{(\ell)}_i = \ell\,\lambda_i \,K_\varepsilon^{(1)} \bigl( \delta_{i}^{(\ell)} \bigr)\,,
\end{equation}
where $K_\varepsilon^{(1)}$ extracts the simple pole in the $\ell$-loop contribution to the renormalisation factor $Z_i$. The tree level contribution is given by the dimension of the coupling such that $\beta^{(0)}_1=-\frac12\varepsilon\lambda_3$ and $\beta^{(0)}_2=-\varepsilon \lambda_4$. The anomalous dimension is also given by 
\begin{equation}\label{eq:fullAnomDim}
    \gamma_\sigma =  \frac12 \sum_{i=3,4} \beta_i \frac{\partial}{\partial \lambda_i} \log Z_{\sigma} \,,
\end{equation}
which can likewise be shown to pick out the simple pole in the field renormalisation counterterm
\begin{equation}\label{eq:simpleAnomDim}
    \gamma^{(\ell)}_\sigma = - \frac12 \ell \, K_\varepsilon^{(1)} \bigl( \delta_{\sigma}^{(\ell)} \bigr) \,.
\end{equation}

\section{Structural theorem and non-renormalisation results}\label{sec:Non-renormalisation theorem}
The \emph{naive} superficial degree of divergence (SDD), $\omega'(\Gamma_n)$, for any amputated $n$-point
graph in this theory is four. To show this we start with 
\begin{equation}
    \omega'(\Gamma_n)= 4\ell-4E+4(n_3+n_4)\,,
\end{equation}
where the first term counts the scaling of the loop momentum measure, the second subtracts the scaling of the $E$ quartic propagators, and the last term adds the scaling of the interaction vertices. Now using the well known Euler's identity for graphs:
\begin{equation}\label{eq:Eulers_thrm}
    \ell-E+V=1 \,,
\end{equation}
with $V=n_3+n_4$ the number of vertices, one easily derives
\begin{equation}
\label{eq:naivesdd}
    \omega'(\Gamma_n)= 4(\ell-E+V)=4\,.
\end{equation}
By itself this result would be troubling since this would mean that all correlators of the theory would be divergent and one would require an infinite number of counterterms rendering this theory non-renormalisable and physically worthless. However, this conclusion would be wrong as we shall show in the following. In fact the numerator of the loop integrand can be factorised  
into two pieces, one of which depends only on the external momenta and thus factors out of the loop integrals. Effectively this reduces the actual degree of the divergence, such that correlators with more than 4 external legs are superficially finite. In the following we shall thus focus on this \textit{effective degree of divergence} which we denote by $\omega(\Gamma_n)$, and elucidate its all-loop consequences.

For a given 1PI diagram we define $v_{i,j}$ to denote the number of $i$-vertices with $j$ external edges. For example, $v_{3,1}$ is the number of external cubic vertices, i.e. those in which one leg can be external, and $v_{3,0}$ is the number of internal couplings, where all legs are internal. Note, since we are concerned with 1PI graphs for renormalisation, this automatically implies that $v_{3,2}=v_{4,3}=0$. Therefore, for any connected 1PI graph, $n_i=\sum_{j=0}^{i-2} v_{i,j}$. We can then write an equation for the number of external legs, $n$, of an $n$-point graph as 
\begin{equation}
\label{eq:nrel}
    n =v_{3,1} + v_{4,1}+ 2v_{4,2}\;.
\end{equation}
To see whether a diagram is finite or divergent we can count the SDD. To do this we need to see what power of external momenta each vertex gives to a diagram. For the quartic interaction this is straightforward, with one external leg we have only one power of external momenta and for two external legs we receive two powers of external momenta as follows from \cref{eq:V4}.

Instead the cubic vertex, since it can be written as a Gram determinant, as follows from \cref{eq:V3_Gram}, scales quadratically with the momenta from each leg. This allows us to derive that the effective degree of divergence for a general $n$-point 1PI graph is given by
\begin{equation}
\label{eq:omega}
    \omega(\Gamma_n) =  \omega'(\Gamma_n) - 2v_{3,1} - v_{4,1} - 2v_{4,2} \,. 
\end{equation}
We see that for the number of external edges for the cubic vertex, $v_{3,1}$, we have included a factor of two. If the effective degree of divergence is less than zero the graph is not superficially divergent, and will therefore not require a local counterterm. 
From \cref{eq:omega} and \cref{eq:nrel} it also follows that
\begin{equation}\label{eq:omega2}
    \omega(\Gamma_n) = 4 - n - v_{3,1} \,. 
\end{equation}
Since $v_{3,1}\ge 0$ we immediately get that one requires $n\le 4$ for there to be an overall divergence at all. 

For the case where $n\le 4$ we now derive a bound on the powers $n_3$ and $n_4$ of the couplings which enter the local counterterm and which eventually determine the beta functions of the theory. Again, for there to be a possible UV divergence we require $\omega\left(\Gamma_n\right)\geq0$, therefore, it follows from \cref{eq:omega2} that 
\begin{equation} 
\label{eq:IRwbound}
    4 - 2n + n_4 - v_{4,0} + v_{4,2}  \ge 0  \,,
\end{equation}
where we have used \cref{eq:nrel} and $n_4 =  v_{4,0} + v_{4,1} + v_{4,2} $. From the strong bounds, $n_4 \geq (n_4 -v_{4,0}) \geq v_{4,2}$, we are able to garner a weak bound, $n_4 \geq (n_4 -v_{4,0}+v_{4,2})/2$, which upon substitution into \cref{eq:IRwbound}, gives the following condition for a UV divergence to be present. Namely, that
\begin{equation}
    n_4  \ge  n - 2 \,.
\end{equation}
Using \cref{eq:numberoflegsforloops} one can also find the following bound for $n_3$ to be $2(\ell+1)-n  \ge n_3 $.
We summarize these statements in the following \emph{structural theorem} about the counterterms of this theory.

\begin{tcolorbox}[ colframe=black, colback=blue!1!white, sharp corners, boxsep=4pt , boxrule=0.3mm]
\textbf{Structural Theorem:}
\emph{The renormalisation factors of the dimension-zero shift invariant theory are constrained by the following  general form}
\begin{align}
    \label{eq:nonrenormphi}
    Z_{\sigma}&=1+\sum_{\ell=1}^\infty\sum_{n_4=0}^{\ell} \left(\lambda_3\right)^{2(\ell-n_4)}\left(\lambda_4\right)^{n_4}\,Z_\sigma^{(2(\ell-n_4),n_4)},
    \\
    \label{eq:nonrenorm1}
    Z_{1}&=1+\sum_{\ell=1}^\infty\sum_{n_4=1}^{\ell} \left(\lambda_3\right)^{2(\ell-n_4)}\left(\lambda_4\right)^{n_4}\,Z_1^{(2(\ell-n_4)+1,n_4)},
    \\
    \label{eq:nonrenorm2}
    Z_{2}&=1+\sum_{\ell=1}^\infty\sum_{n_4=2}^{\ell+1} \left(\lambda_3\right)^{2(\ell-n_4+1)}\left(\lambda_4\right)^{n_4-1}\,Z_2^{(2(\ell-n_4+1),n_4)}.
\end{align}
\end{tcolorbox}

Safari {\it et al.} considered kinetic terms $\sigma \Box^k \sigma$ with $k\geq 2$, with the same shift-invariant interactions studied here. They treated the cubic and quartic interactions as separate cases, the former being odd in $\sigma$ and the latter even. 
Using heat kernel methods, they found $Z_1=1$ at one loop due to non-trivial cancellations in the effective action and speculated that this might be due to a hidden supersymmetry. Using diagrammatic methods, we are able to show that $Z_1=1$ to all orders, as a consequence of the Gramian nature of the cubic vertex. Thus, when the quartic coupling $\lambda_4=0$, we obtain an all-loop Ward-like identity,
\begin{equation}\label{eq:secondWI}
    Z_1 = Z_3 Z_\sigma^{3/2} = 1 \,.
\end{equation}
In this case, for a purely cubic interaction, the action is invariant under {\it Galileon} symmetry \cite{Nicolis:2008in,Kampf:2014rka}: 
 \begin{equation}\label{eq:Galileon}
    \delta \phi= b_\mu x^\mu \rightarrow \delta \partial_\mu \phi= b_\mu,
\end{equation}
with $b^\mu$ any constant four-vector, as is easily checked using integrations by parts. Our results show that this symmetry is non-anomalous: it may provide a clue as to why the purely cubic interaction is RG invariant. Note that Galileon symmetry \cref{eq:Galileon} has previously been observed to imply the non-renormalisation of low order operators in effective field theories of the Galileon type~\cite{Luty:2003vm,Hinterbichler:2015pqa,Goon_2016}. Similarly, improved IR behaviour has been previously seen as a consequence of Gramian-type interactions in other effective field theories including DBI models and nonlinear sigma models~\cite{Cheung:2014dqa,Cheung:2015ota}. We now turn to a discussion of IR divergences in the family of UV-complete theories studied here. 

\section{Euclidean IR finiteness}\label{sec:IRfiniteness}
A major concern for higher-derivative theories is the possible presence of Euclidean IR divergences\footnote{We call these Euclidean IR divergences since this class describes all IR divergences which could be present in non-exceptional, Euclidean kinematics. However this class may of course also be present in non-exceptional off-shell Minkowskian kinematics, covering all IR divergences in this regime too. More complicated Minkowskian IR divergences, including collinear or mass divergences, generally only show up in the on-shell limit.}  due to the increased powers of momenta in the propagators. We show that by including the cubic and quartic shift-invariant interactions we are able to tame these divergences and in fact we have IR finiteness to all orders in the $n$-vertices. To formulate our proof we employ the terminology of refs.~\cite{Brown:2015fyf,Beekveldt:2020kzk}:
\begin{itemize}
    \item The term \emph{mass-momentum spanning} (m.m.) means that the subgraph $\gamma$ contains all the external momenta and internal masses of its parent graph $G$. 
    \item A subgraph $\gamma$ is \emph{motic} if $\gamma_0$, formed when contracting all the external lines of $\gamma$ into an auxiliary vertex $v_0$, is 1PI.  
\end{itemize}

For a general Feynman graph, $G$, we can then isolate possible IR divergent subgraphs by considering the following IR factorization: 
\begin{equation}\label{eq:softfactorization}
    G \sim T_{p_s}^{(n_\mathrm{min})}(\gamma) \cdot G/\gamma \,,
\end{equation}
where $\gamma$ are m.m. and motic. The IR divergent part is instead captured by the contracted graph $G/\gamma$ which is obtained by contracting $\gamma$ into a point in $G$. Here 
$ T_{p_s}^{(n_\mathrm{min})}$ denotes Taylor expansion in the IR momenta to leading order, which we denote by $n_\mathrm{min}$ to signify that it may (and in fact will) not be zero. 

One further decomposes $G/\gamma$ into its different 1VI\footnote{1VI stands for one-vertex-irreducible, which is a graph which can not be disconnected through the removal of any one vertex.} components, i.e. $G/\gamma=\sqcup_i \tilde\gamma_i$. The $\tilde\gamma_i$ are then called IR-irreducible and must have non-positive SDD for this to lead to an IR divergence. We illustrate this decomposition diagrammatically as follows:
\begin{equation}
\label{eq:IRgraph}
G =    \GGraph\quad ,\quad  \;\;\;\; G/\gamma = \;   \Gsubgammagraph \,, \qquad  \gamma = \subgammagraph \quad \,.
\end{equation}
We see that the vertex $v_\gamma$ in $G/\gamma $ arises from the contraction of $\gamma$ into a point. Note that this vertex will also carry momentum dependence stemming from the Taylor expansion $\gamma$ around the infrared limit of the lines $k_1,\dots,k_m \to 0$. What we wish to show is that at the leading order in this Taylor expansion we do not have an IR divergence. For $G$ to be potentially IR divergent we require that there exists at least one $\gamma$ such that $G/\gamma$ has SDD\footnote{Note that since $G/\gamma$ is a vacuum graph its naive SDD  $\omega'$ and its effective SDD $\omega$ coincide.} less than or equal to zero, that is
\begin{equation}
    {\omega}\left( G/\gamma \right) \leq 0 \, \quad \text{for some } \gamma \subset G\,.
\end{equation}
In \cref{eq:IRgraph} we further decompose the IR graph whose edges and vertices make up $G/\gamma$ into the edges $k_i$, which connect to $\gamma$, and $\hat\gamma$, which we assume to be a general connected graph.  However, this IR divergence can still be tamed if the leading degree in the IR momenta in $\gamma$ is sufficiently greater than zero, leading to additional IR scaling of $v_\gamma$. This is indeed always the case for the theory under consideration, since each interaction vertex Feynman rule has at least linear scaling with each momentum of its adjacent lines. This therefore motivates defining the quantity
\begin{equation}
{\omega}_{\mathrm{IR}}^{\mathrm{min}}(G/\gamma)=\omega(G/\gamma)+\omega_{\mathrm{IR}}(v_\gamma)\,,
\end{equation}
 which we call the effective IR SDD, where $\omega_{\mathrm{IR}}(v_\gamma)$ is the IR scaling degree of $v_\gamma$. From the graphs in  \cref{eq:IRgraph} we then deduce the relation:
\begin{equation}\label{eq:IRdivergence}
    {\omega}_{\mathrm{IR}}^{\mathrm{min}}\left(  G/\gamma \right) = {\omega}'\left( \hat{\gamma} \right) + 4(m-1) - 4m + \omega_{\mathrm{IR}}\left( v_\gamma \right) \,.
\end{equation}
This is derived by counting the powers of momenta present in the diagram. 

First we must include the naive SDD for the subgraph $\hat{\gamma}$, which for a general graph in this theory was given in \cref{eq:naivesdd} to be ${\omega}'\left( \hat{\gamma} \right) = 4$. Note that we use the naive SDD here since the external momenta of $\hat \gamma$ are still part of the IR subgraph. We receive a further $m-1$ loops from the edges $k_1,\dots,k_m$, leading to $4(m-1)$ from the measure and also four inverse powers of momenta for each of the $m$ legs $k_i$. Finally, we must include the contribution of the vertex $v_\gamma$ which arises from the contraction of the graph $\gamma$. Usually in a theory with no derivative interactions this is trivially zero, however in our case we see that in the minimal case each leg $k_i$ scales with at least one power of momenta, therefore $\omega_{\mathrm{IR}}\left(v_{\gamma} \right) \geq m$. Putting all of this together we find 
\begin{equation}\label{eq:IRfinite}
   {\omega}_{\mathrm{IR}}^{\mathrm{min}}\left(  G/\gamma \right) \geq m \,.
\end{equation}
For $m \geq 2$ it thus follows that there are no IR divergences in the off-shell-Feynman diagrams in this theory.

Of course, one could consider the case $m=1$. This would either correspond to a tadpole, which would make the diagram scaleless and therefore we need not discuss it, or, otherwise, we could have the case where $\hat\gamma$ is trivial in the sense that it has no interaction vertex so that the line $k_1$ passes straight through and out as $k_2$. A similar argument as before also holds for this case since there are $4$ inverse powers from the single IR line, 4 powers of loop momentum measure and at least 2 powers from $v_\gamma$, leading to ${\omega}_{\mathrm{IR}}^{\mathrm{min}} \ge 2$.

Therefore, we have shown that the off-shell correlation functions in the theory are IR finite to all orders in perturbation theory.

\subsection{Double-bubble example}

We consider a well-known example of an IR divergent two loop graph for $\phi^3$ theory \cite{Larin:2002sc}. As we shall see, the same graph is IR finite for the theory we are concerned with. This graph is given as follows.
    \begin{equation}
        \BubbleGraph  = G_{\phi^3} = \int d^4k_1d^4k_2 \frac{1}{(k_1^2)^2} \frac{1}{(p-k_1)^2} \frac{1}{k_2^2} \frac{1}{(k_1 - k_2)^2} \,.
    \end{equation}
In the soft limit of $k_1,k_2\to 0$ the graph factorises as
\begin{equation}
    G_{\phi^3}  \sim \frac{1}{p^2}\int d^4k_1d^4k_2 \frac{1}{(k_1^2)^2}  \frac{1}{k_2^2} \frac{1}{(k_1 - k_2)^2} \,.
\end{equation}
We can identify the following diagrams according to \cref{eq:softfactorization} which respectively correspond to the motic m.m. and IR subgraphs 
\begin{equation}
   \gamma =  \GamEx \quad = \frac{1}{p^2}  \,, \qquad  G_{\phi^3}/\gamma =   \GsubGEx  = \int d^4k_1 \, d^4k_2  \frac{1}{(k_1^2)^2} \frac{1}{k_2^2} \frac{1}{(k_1 - k_2)^2}  \,.
\end{equation}
In $\phi^3$ theory an IR divergence has to satisfy the bound 
\begin{equation}
    {\omega}_{\mathrm{IR}}^{\mathrm{min}} \left( G/\gamma  \right) = {\omega}'(\hat{\gamma}) + 4(m-1) - 2m + \omega_{\mathrm{IR}}(v_\gamma) \leq 0 \,,
\end{equation}
where we have adjusted \cref{eq:IRdivergence} to reflect the different quadratic propagator power. For this diagram $m=2$ and ${\omega}'(\hat{\gamma}) = 0$ since the bubble subgraph is log-divergent, similarly the vertex $v_\gamma$ does not scale with $k_1$ or $k_2$ and so $\omega_{\mathrm{IR}}(v_\gamma)=0$. Altogether for this example we find
\begin{equation}
  {\omega}_{\mathrm{IR}}^{\mathrm{min}}\left(  \GsubGEx \right) = 0  \,,
\end{equation}
and therefore $G_{\phi^3}$ has a logarithmic IR divergent subgraph (which indeed leads to a $1/\epsilon$ pole of IR origin).  

Instead, for the shift symmetric four-derivative theory we have, for the same diagram, the integral representation,
\begin{equation}
        G_{\sigma} = \int d^4k_1d^4k_2 \frac{V_3(p,k_1,-k_1-p)^2V_3(k_1,k_2,-k_1-k_2)^2}{(k_2^4)(k_1^4)^2(p-k_1)^4(k_1 - k_2)^4} \,.
    \end{equation}
In the IR limit $k_1,k_2\to 0$ we then get the following factorisation (we only indicate the overall scaling of the momenta)
\begin{equation}
   \gamma \sim \frac{p^4}{p^4} \,, \qquad  \qquad \qquad G_{\sigma}/\gamma \sim \int d^4k_1 \, d^4k_2  \frac{k_2^4k_1^8}{(k_2^4)(k_1^4)^2(k_1 - k_2)^4}  \,.
\end{equation}
We see in agreement with \cref{eq:IRfinite} that we find
\begin{equation}
  {\omega}_{\mathrm{IR}}^{\mathrm{min}}\left(  G_{\sigma}/\gamma  \right) = 4 \,,
\end{equation}
where $\omega_{\mathrm{IR}}(v_\gamma) = 4$ and ${\omega}'(\hat{\gamma}) = 4$. In this example, because of the numerous cubic vertices, we find the best possible behaviour for $\omega_{\mathrm{IR}}(v_\gamma)$, so the degree of divergence is much higher than the minimal case of $m=2$  which would yield ${\omega}_{\mathrm{IR}}^{\mathrm{min}}\left(  G_{\sigma}/\gamma  \right)\ge 2$ here. 

\section{The perfect square theory and its \texorpdfstring{$\phi^4$}{phi 4} counterpart}\label{sec:Oclassicaly}

In his analysis of the RG coupling flow in these theories, Holdom observed that a particular trajectory in coupling space, $\lambda_4 = -\frac12 \lambda_3^2$, is RG invariant at one loop. Writing $\lambda_4=-\lambda^2/2$ and $\lambda_3=-\lambda$, the Lagrangian density \cref{eq:LagArbCoupl} along this trajectory takes the form of a perfect square: 
\begin{equation}
\label{eq:perfectsquare}
    \mathcal{L}_{PS} = - \frac12( \Box \sigma + \lambda (\partial_\mu \sigma \partial^\mu \sigma))^2\;.
\end{equation}
Holdom also found that, on this special trajectory, the differential cross section is positive and independent of the scattering angle, reminiscent of scattering in a scalar theory with non-derivative  interactions~\cite{Holdom:2023usn,Holdom:2024cfq}.  Recently, for this special perfect square theory, two of us generalised Holdom's results to arbitrary covariant transition probabilities~\cite{Bateman:2026letter}, and explained the positivity of transition probabilities as being due to a hidden ghost parity symmetry. That argument followed from an embedding of the perfect square theory into a two-derivative $O(1,1)$ model. This embedding allows the running of the coupling along this perfect square trajectory to be related to that of a simple $\phi^4$ scalar.

To see this, we consider the well known $O(2)$-invariant $\phi^4$ scalar theory described by the Lagrangian
\begin{equation}\label{eq:phi4}
	\mathcal{L}_{\phi^4}= \frac{1}{2} \partial_\mu \phi_a\,\partial^\mu \phi_a - \frac{g}{4!} (\phi_a \phi_a)^2 \,,
\end{equation}
where the index $a=1,2$. Note that the beta function for the coupling $g$ is independent of the metric on field space. Consequently, the beta function for the $O(2)$ model is equivalent to that of the $O(1,1)$ model described by the Lagrangian
\begin{equation}\label{eq:OY}
    \mathcal{L}_{\Omega \Upsilon} = \partial_\mu \Omega \, \partial^\mu \Upsilon - \frac{g}{6} (\Omega \Upsilon )^2.
\end{equation}
To relate these two Lagrangians, we take $\phi_2$ to be imaginary and consider the field transformation $\phi_1=(\Omega+\Upsilon)/\sqrt{2}$ and $\phi_2=i(\Omega-\Upsilon)/\sqrt{2}$, such that the two real fields $\Omega$ and $\Upsilon$ are the lightcone coordinates on the Lorentzian plane. To relate the $O(1,1)$ invariant theory to the perfect square, we restrict to a sub-sector of the theory in which $\Omega$ is strictly positive. The field $\Upsilon$ can then be integrated out exactly and we find that
\begin{equation}
\label{eq:pathintegralactions}
    \int_{\Omega>0}\mathcal{D}\Omega\,\mathcal{D}\Upsilon \,e^{i\int d^4x \mathcal{L}_{\Omega\Upsilon}} = c \int_{\Omega>0}\frac{\mathcal{D}\Omega}{\Omega}\, e^{\frac{i}{2g}\int d^4x \left(\frac{\square\Omega}{\Omega}\right)^2} = c' \int \mathcal{D}\sigma \, e^{i \int d^4x \mathcal{L}_{PS}} \,,
\end{equation}
where $c$ and $c'$ are constants, and the final equality follows from the field redefinition 
\begin{equation}\label{eq:fieldrefefiniton}
    \Omega=e^{\lambda\sigma} \,.
\end{equation} The shift symmetry of the perfect square theory corresponds to the scale symmetry of the $O(1,1)$ theory, which is the connected $SO^\uparrow(1,1)$ subgroup. The two couplings are related by
\begin{eqnarray}
    \lambda^2 = -\frac{g}{3}\,,
\end{eqnarray}
which means that we must consider the $O(1,1)$ model at negative coupling. This ensures that the Euclidean action in the middle term in \cref{eq:pathintegralactions} is positive. This path integral derivation implies that the beta function along the perfect square trajectory is related to that of the $O(2)$-invariant $\phi^4$ theory at negative coupling, that is,
\begin{equation}\label{eq:equivalencebetarelationship}
   \left. \beta_\lambda = - \frac{1}{6\lambda}\beta_g \right|_{g = - 3\lambda^2 } \;.
\end{equation}
The relative minus sign is responsible for the asymptotic freedom of the perfect square theory. This beta function equivalence holds to all orders. In Section \ref{sec:betafunsequivalence} we verify this agreement explicitly up to three loops.

In a companion paper we show that the perfect square theory, and the $O(1,1)$-symmetric model \cref{eq:OY} it is embedded in, may be used to describe quadratic gravity (QG) in a certain limit of coupling constants where the only interacting degree of freedom is the local scale factor of the metric~\cite{Anderson:2026Rsqaured}. This is the conformally flat limit of quadratic gravity (CFQG). The diffeomorphism invariance of QG and its conformally flat limit (CFQG) yields a Ward identity which ensures that the only local term allowed in the action is the integral of the four-dimensional Ricci scalar squared. The Ward identity is responsible for protecting the perfect square form of the Lagrangian density under renormalisation. It implies the identities 
   \begin{equation}\label{eq:Warditentity}
    Z_3Z_\sigma^{1/2} = Z_4 Z_\sigma\,=1.
\end{equation}
which were observed by Holdom at one loop \cite{Holdom:2023usn} and which  we confirm, in section \ref{sec:specialline}, up to three loops. We find it fascinating that a {\it gravitational} Ward identity can protect a basic structural property of a theory defined in flat spacetime, even though that theory does not involve gravity. It would be interesting to find other examples. 

\section{Computations and Results}
\label{sec:ComputationsandResults}
\subsection{\texorpdfstring{$R^*$}{R*} Method} To compute the renormalisation constants  we make use of a private implementation, $R$-graph, of the $R^*$-method 
\cite{Chetyrkin:1982nn,Vladimirov:1979zm,
Chetyrkin:1984xa,Smirnov:1985yck,Larin:2002sc,Kleinert:2001hn,Chetyrkin:2017ppe,Herzog:2017bjx,deVries:2019nsu,Beekveldt:2020kzk,Henriksson:2025hwi} in the $\overline{MS}$ scheme. The program is implemented in Maple \cite{maple} and FORM \cite{Kuipers:2012rf,Vermaseren:2000nd,Ruijl:2017dtg} and makes use of the FORM libraries FORCER \cite{Ruijl:2017cxj} and OPITeR \cite{Goode:2024cfy}. The Feynman diagrams are generated by \texttt{QGRAF} \cite{Nogueira:1991ex}. 

The main advantage of the $R^*$ method is that it allows one to extract the local UV counterterm for a given $n$-vertex $\Gamma$ directly in a fully automated manner. This makes it a convenient tool for computations. A naive extraction of such counterterms would require the computation of $n$-vertices directly, however in the $R^*$ method the divergences are extracted from products of 2-point functions each of at most $\ell-1$ loops, thus radically simplifying the complexity of the individual integrals to be computed. To compute our results we also made use of a recently developed $R^*_{\textrm{ME}}$ operation, which is more efficient than the standard diagrammatic $R^*$ operation. It was recently employed in ref.~\cite{Henriksson:2025hwi} for the computation of anomalous dimensions of higher dimensional operators in scalar EFTs, the basic idea was introduced in Chakraborty's master's thesis \cite{Chakraborty2023}.

The key idea of the $R^*$ method is that the overall or local UV counter term of a given graph only depends polynomially on the external momenta. The coefficients of different monomials of the counterterms can thus be projected out by acting with suitable differentiation operators on the correlators. After this one can, in principle, set the external momenta to zero since they no longer affect the value of the counter term. This is however spoiled if the  nullification of external momenta produces IR divergences, as it often does. The $R^*$ method deals with this problem by including suitable IR counterterms. The problem with this is that these counterterms can lead to complicated recurrences which can lead to expensive combinatoric overhead, especially if the number of derivatives is high. For the high-derivative theory under consideration this is indeed a problem since all the $n$-vertices have quartic scaling in the external momenta. 

$R^*_{\textrm{ME}}$ deals with the problem more efficiently in that it subtracts the divergences created in the Taylor expansion via a small momentum asymptotic expansion which is a non-recursive procedure. We refer to ref.~\cite{Henriksson:2025hwi} for more details and explicit examples. However, even with $R^*_{\textrm{ME}}$ it was difficult to compute the renormalisation counterterm of the $3$-loop 4-vertex directly. To accomplish this particular calculation we therefore employed a method solely based on the momentum expansion
by subgraph \cite{Chetyrkin:1982zq,Chetyrkin:1983qlc,Gorishnii:1983su,Gorishnii:1986gn,Chetyrkin:1988zz,Chetyrkin:1988cu,LlewellynSmith:1987jx,Gorishnii:1989dd,Smirnov:1990rz,Smirnov:1994tg,Smirnov:2002pj,Chakraborty:2024uzz}. This method was also employed to compute the anomalous dimensions of alien operators of spin-$N$ twist-2 operators in QCD \cite{VanThurenhout:2025vui,Falcioni:2024xav,Falcioni:2023luc,Falcioni:2023vqq,Falcioni:2024xyt,Falcioni:2024qpd}. 

Let $T^{(m)}_{\{p\}}$ denote the \emph{Taylor-expansion-operator} projecting out the order $m$ term of the series expansion in the set of external momenta $\{p\}$. Importantly, $T^{(m)}_{\{p\}}$ does not commute with integration, since there are potential singularities in the limit $p\to 0 $, which effectively render the series into a log-power series in scalar products of the momenta $\{p\}$. 
An operator which does commute with integration is the asymptotic expansion operator $\widetilde {T}^{(m)}_{\{p\}}$ and thus it can act on the integrand before integration. It is defined via the expansion-by-subgraph as follows:
\begin{equation}
\widetilde {T}^{(m)}_{\{p\}}(G)=\sum_{\gamma_A \subset G} T^{(m)}_{\{p\}}(\gamma_A)* G/\gamma_A\,,
\end{equation}
where $\gamma_A$ is an asymptotically irreducible subgraph; see ref.~\cite{Smirnov:2002pj,Chakraborty:2024uzz} for more details. 

The finiteness of $\Gamma_{n}$ under renormalisation was expressed in \eqref{eq:deltaZ1} and \eqref{eq:deltaZ2}. Rearranging these equations and expanding, around a suitable set of external momenta $\{p\}$, on the left and right hand sides, to a suitable order $m$, we then obtain
\begin{align}
  \lambda_3 T^{(m)}_{\{p\}}(V_3)  (Z_1-1) = - K_\varepsilon \left( Z^{3/2}_\sigma \widetilde T^{(m)}_{\{p\}}\delta \Gamma_3^r(Z_3\lambda_3,Z_4\lambda_4)\right)\,,\\
\lambda_4 T^{(m)}_{\{p\}}(V_4)  (Z_2-1) = -K_\varepsilon \left( Z^{2}_\sigma \widetilde T^{(m)}_{\{p\}} \delta \Gamma_4^r(Z_3\lambda_3,Z_4\lambda_4)\right)\,.
\end{align}
To choose a suitable $\{p\}$ and $m$ one chooses $n-2$ independent external momenta of the correlator, such that one still has one momentum left providing scale to the integrals. The order $m$ should be so as to annihilate all dependence on the external momenta, e.g.
\begin{itemize}
    \item $m=2$ with $\{p_3\}$ for $\Gamma_3^r$, 
    \item $m=2$ with $\{p_3,p_4\}$ for $\Gamma_4^r$.
\end{itemize}
For $\Gamma_2^r$ there is no advantage to be gained from this method, since one already only needs to compute 2-point functions. We have, nevertheless, performed this calculation as well, confirming the $R^*_\textrm{ME}$ result. 

Regarding computational cost the most complicated calculation was the momentum expansion of the 4-vertex, which took about 2 days on a desktop computer to compute all 9052 diagrams. With $R^*_{\textrm{ME}}$ it would have likely taken at least a week without further optimisation.

\subsection{3 loop results}

The calculated counterterms up to 3 loops in $D = 4 - \varepsilon$ are 
\begin{align}
    \delta_\sigma^{(1)} =& \frac{10}{\varepsilon}\lambda_3^2 \,,\\
    \delta_\sigma^{(2)} =& \left(\frac{530}{9\varepsilon}   - \frac{100}{\varepsilon^2}\right)\lambda_3^4 + \left(\frac{520}{9 \varepsilon } - \frac{200}{\varepsilon^2} \right)\lambda_3^2\lambda_4 \,, \label{eq:deltaPhiL2} \\
    \delta_\sigma^{(3)} =& \left(\frac{544}{3\varepsilon}  \zeta_3 + \frac{9766}{9\varepsilon}- \frac{5300}{3\varepsilon^2}  + \frac{5000}{3\varepsilon^3} \right)\lambda_3^6  + \left(\frac{128}{\varepsilon}  \zeta_3 + \frac{525352}{243\varepsilon}- \frac{114400}{27\varepsilon^2} \right. \nonumber\\
    & \left. + \frac{16000}{3\varepsilon^3} \right)\lambda_3^4\lambda_4  + \left(\frac{128}{3\varepsilon}  \zeta_3 + \frac{208844}{243\varepsilon}- \frac{48800}{27\varepsilon^2} + \frac{4000}{3\varepsilon^3} \right)\lambda_3^2\lambda^2_4  + \frac{320}{3\varepsilon}\lambda_4^3 \,,\\    
&\nonumber\\
    \delta_1^{(1)} =& -\frac{20}{\varepsilon}\lambda_4 \,, \\
    \delta_1^{(2)} =& - \left( \frac{760}{9 \varepsilon} - \frac{200}{\varepsilon^2}\right)\lambda_3^2 \lambda_4 - \left(\frac{440}{9\varepsilon} - \frac{400}{\varepsilon^2} \right)\lambda_4^2 \,, \\
    \delta_1^{(3)} =& - \left(\frac{1088}{3\varepsilon}  \zeta_3 + \frac{47696}{27\varepsilon} - \frac{26800}{9\varepsilon^2} + \frac{10000}{3\varepsilon^3} \right)\lambda_3^4\lambda_4    - \left(\frac{256}{\varepsilon}  \zeta_3 + \frac{761984}{243\varepsilon}\right.\nonumber \\
    &\left. - \frac{171200}{27\varepsilon^2}  + \frac{32000}{3\varepsilon^3} \right)\lambda_3^2\lambda^2_4  - \left(\frac{256}{3\varepsilon}  \zeta_3 + \frac{206728}{243\varepsilon}
    - \frac{42400}{27\varepsilon^2}  + \frac{8000}{\varepsilon^3} \right)\lambda_4^3\,, \\
&\nonumber\\
    \delta_2^{(1)} =& -\frac{20}{\varepsilon}\lambda_4 \,,\\
    \delta_2^{(2)} =& - \left(\frac{520}{9\varepsilon} - \frac{200}{\varepsilon^2}\right)\lambda_3^2\lambda_4 + \left(\frac{40}{9\varepsilon} + \frac{400}{\varepsilon^2} \right)\lambda_4^2 \,, \\
    \delta_2^{(3)} =& - \left(\frac{1088}{3\varepsilon}  \zeta_3 + \frac{12592}{9\varepsilon} - \frac{7600}{3\varepsilon^2} + \frac{10000}{3\varepsilon^3} \right)\lambda_3^4\lambda_4  - \left(\frac{256}{\varepsilon}  \zeta_3 + \frac{527504}{243\varepsilon} \right.\nonumber \\
    &\left.- \frac{123200}{27\varepsilon^2} + \frac{32000}{3\varepsilon^3} \right)\lambda_3^2\lambda^2_4 - \left(\frac{256}{3\varepsilon}  \zeta_3 + \frac{94888}{243\varepsilon} - \frac{5600}{27\varepsilon^2}  + \frac{8000}{\varepsilon^3} \right)\lambda_4^3 \,.
\end{align}
We remark that at three loops there is an appearance of Riemann's  $\zeta_3\approx 1.202056903$, this is to be expected from a three loop renormalisation group calculation. Moreover, terms including $\pi^2$  cancel out as expected \cite{Boito:2016pwf, Davies:2017hyl, Baikov:2018wgs}. To guarantee the correctness of the expressions we have performed several checks.  Our one loop results are in agreement with the results in \cite{Holdom:2023usn,Holdom:2024cfq} and the overlapping cases from \cite{Safari:2021ocb,Tseytlin:2022flu}. Furthermore, we have confirmed the absence of logarithms of kinematic scales, which appear at intermediate steps in the calculation, but must finally cancel in the renormalisation constants.  Further consistency checks for the \emph{perfect square} theory are reported below.  

We now list some more observations about the structure of the renormalisation constants, and their consistency with the consequences of the structural theorem 
\cref{eq:nonrenormphi,eq:nonrenorm1,eq:nonrenorm2}.
For $\delta_\sigma$ we observe:
\begin{itemize}
    \item No $\lambda_4$ at one loop, this is because the only contributing diagram is scaleless. This does not follow from our theorem. 
    \item No $\lambda_4^2$ at two loops. There is one non-vanishing diagram, but it yields no contribution to the counterterm. This does not follow from our theorem. 
    \item At three loops all coupling combinations are present in accord with \cref{eq:nonrenormphi}.
\end{itemize}
Notably, the first non-vanishing term for contributions coming only from quartic vertices appears transcendentally suppressed, contributing only through a simple $1/\epsilon$ pole. Understanding the underlying mechanism for this suppression would be interesting, especially since this is not addressed by the structural theorem. 

For $\delta_1$ we observe no contributions scaling with $\lambda_3^{2\ell}$ at $\ell$ loops. This is in accordance with \cref{eq:nonrenorm1}. For $\delta_2$ we observe no contributions scaling with $\lambda_3^{2\ell+2}$ or  $\lambda_3^{2\ell}$ at $\ell$ loops, again in accordance with \cref{eq:nonrenorm2}. 

Using these results we can find the three loop beta functions to be 
\begin{align}\label{eq:Beta3ThreeLoops}
    \beta_3 =& -\frac{\varepsilon}{2}\lambda_3 - \left[ 20\lambda_3 \lambda_4 + 15\lambda_3^3 \right]  - \left [ \frac{880}{9}\lambda_3\lambda_4^2+ \frac{3080}{9}\lambda_3^3\lambda_4  + \frac{530}{3}\lambda_3^5 \right] \nonumber\\
    & - \left[ \left(  \frac{245608}{81} + 256\zeta_3\right)\lambda_3\lambda_4^3 + \left(  \frac{1075250}{81} + 960\zeta_3\right)\lambda_3^3\lambda_4^2 \right.\nonumber \\
    &\left.+ \left(  \frac{405764}{27} + 1664\zeta_3\right)\lambda_3^5\lambda_4  + \left( 816\zeta_3 + 4883\right)\lambda_3^7  \right] + \cdots \;,
\end{align}
\begin{align}
\label{eq:Beta4Threeloops}
    \beta_4 =& -\varepsilon \lambda_4 - \left[20\lambda_3^2 \lambda_4 + 20\lambda_4^2 \right] - \left[  \frac{2120}{9}\lambda_3^4\lambda_4  + \frac{1040}{3}\lambda_3^2\lambda_4^2 + \frac{80}{9}\lambda_4^3 \right]\nonumber  \\
   & -\left[  \left( \frac{19532}{3} + 1088\zeta_3\right)\lambda_3^6\lambda_4  + \left( \frac{1390688}{81} + 1856 \zeta_3 \right)\lambda_3^4\lambda_4^2      \right. \nonumber \\
   &\left.+ \left( \frac{315064}{27} + 1024\zeta_3\right)\lambda_3^2\lambda_4^3 + \left(\frac{146728}{81} + 256\zeta_3 \right) \lambda_4^4 \right] + \cdots  \;,
\end{align}
where the ellipses denote higher-loop corrections. We observe the expected absence of poles in $\varepsilon$ of the beta functions $\beta_3$ and $\beta_4$. This yields another cross check for the correctness of the expressions. For completeness, we also give the anomalous dimension using \cref{eq:fullAnomDim}: 
\begin{align}
    \gamma_{\sigma} =& -\,5\lambda_3^2 - \left(\frac{530}9\lambda_3^4 + \frac{520}{9}\lambda_3^2\lambda_4\right) - \left( 272\zeta_3 + \frac{4883}{3} \right)\lambda_3^6  - \left( 192\zeta_3 + \frac{262676}{81}\right)\lambda_3^4\lambda_4 \nonumber \\
    & - \left( 64\zeta_3 + \frac{104422}{81} \right) \lambda_3^2\lambda_4^2 - 160\lambda_4^3 + \cdots \,.
\end{align}
This provides further consistency checks since we observe, again, the appropriate cancellation of  poles in $\varepsilon$ and equivalence with \cref{eq:simpleAnomDim}.

\subsection{Perfect square results}
\label{sec:specialline}

As discussed, there exists a special solution to the two beta functions that relates the coupling constants at one loop, which is
\begin{equation}\label{eq:specialsolution}
    \lambda_4 = -\frac12 \lambda_3^2\,.
\end{equation} 
Using this and $\lambda_3 = -\lambda$ we see explicitly from the three loop beta function results, \cref{eq:Beta3ThreeLoops} and \cref{eq:Beta4Threeloops}, that this special solution is preserved. Therefore, for the perfect square theory we find the beta function up to three loops to be 
\begin{equation}\label{eq:BetaThreeloopPS}
    \beta_\lambda = -\frac{\varepsilon}{2}\lambda - 5\lambda^3   - 30\lambda^5 - \left(192 \zeta_3 +  \frac{617}{2}\right)\lambda^7 + \cdots \;.
\end{equation}
Similarly, we find the anomalous dimension to be given by 
\begin{equation}\label{eq:AnomThreeloopPS}
    \gamma_\sigma =  -5\lambda^2   -  30\lambda^4 - \left(192 \zeta_3 +  \frac{617}{2}\right)\lambda^6 + \cdots \;.
\end{equation}

\subsubsection{Ward identity}
We are able to confirm the Ward identity, given by \cref{eq:Warditentity}, also holds to three loops. Similarly, by substituting \cref{eq:specialsolution} and $\lambda_3 = -\lambda$ into our counterterms, we see that
\begin{equation}\label{eq:threeloopresult}
 \delta_\sigma = \delta_1 =\delta_2   = \frac{10}{\varepsilon} \lambda^2 + \frac{30}{\varepsilon} \lambda^4 + \frac{128\zeta_3}{\varepsilon}\lambda^6 + \frac{617}{3\varepsilon}\lambda^6 - \frac{100}{\varepsilon^2}\lambda^6 + \cdots \;.
\end{equation}
This confirms the following relationship between the renormalisation factors
\begin{equation}\label{eq:Zphi=Z_1=Z-2}
    Z_\sigma = Z_1 =Z_2  \;.
\end{equation}
Using our definitions for $Z_1 = Z_\sigma^{3/2}Z_{3}$ and $ Z_2 =Z_\sigma^2 Z_{4}$, we see this relationship means that the following holds up to three loops:
\begin{equation}\label{eq:Z_4 = Z^2_3}
     Z_{4}  =  Z^2_3 = Z_\sigma^{-1} \;.
\end{equation}
This is precisely the Ward identity in \cref{eq:Warditentity}. This means that for the perfect square theory we can deduce the coupling renormalisation from that of the field renormalisation constant. Computationally this is of course a great advantage in contrast to having to compute the full 3-point and 4-point correlation functions. 

Consequently, if we consider the perfect square Lagrangian given by \cref{eq:perfectsquare}, and use the Ward identity above, we find that
\begin{equation}
    \mathcal{L}   = -\frac{Z_\sigma}{2}\left(\Box \sigma + \lambda (\partial_\mu \sigma \partial^\mu \sigma)\right)^2\;.
\end{equation}
If we perform the field redefinition
\begin{equation}
   \sigma^b  = \frac{\tilde{\sigma}^b}{\lambda^b} \,,
\end{equation}
where by definition $\tilde{\sigma}^b = \tilde{\sigma}$, so that, like in Yang-Mills theories our coupling constant can be brought outside the renormalisation invariant term
\begin{equation}
    \mathcal{L}   = - \frac{1}{2Z_\lambda^2\lambda^2}( \Box \tilde{\sigma} +  (\partial_\mu \tilde{\sigma}\partial^\mu \tilde{\sigma}))^2\;.
\end{equation}

We remark further that this situation is completely analogous to the one in gauge theory as long as one works in a background field gauge \cite{Abbott:1981ke}. Unlike that context we do not require gauge fixing terms and thus there is no need for working in a fixed background. We also note that the result shown in \cref{eq:threeloopresult} has another subtle relation to the background field method, namely the cancellation of $\varepsilon$ poles at the same weight as the $\ell$-loop order. For example, at two-loop we see that the $1/\varepsilon^2$ cancel, although these appear for arbitrary couplings as seen for example in \cref{eq:deltaPhiL2}.

\subsubsection{Beta function equivalence}\label{sec:betafunsequivalence}

The embedding detailed in Section \ref{sec:Oclassicaly} showed that the coupling constant renormalisation for the perfect square and $O(1,1)$ theories should agree to all orders. With actions given respectively by \cref{eq:perfectsquare} and \cref{eq:OY} and coupling constants related by 
\begin{equation}\label{eq:gPhi4PSequiv}
    g=-3 \lambda^2\,.
\end{equation}
The renormalisation for the coupling constant, $g$, given by the $O(1,1)$ symmetric theory is structurally identical to an $O(2)$ symmetric theory, whose action is given by \cref{eq:phi4}. Therefore, we can compare the beta function result for the perfect square at three loops, \cref{eq:BetaThreeloopPS}, to the known calculation for an $O(2)$ symmetric theory. While the five-loop results were known already for a long time  \cite{Chetyrkin:1981jq,Gorishnii:1983gp,Kazakov:1983dyk},   
the current state-of-the-art for $O(2)$ is six loops. The result was given for $O(N)$ theories in \cite{Kompaniets:2017yct}, with the corresponding 6 loop field renormalisation computed earlier in ref.~\cite{Batkovich:2016jus}. For real $\phi^4$ the beta function is known to 7 loops \cite{Schnetz:2022nsc,Schnetz:2016fhy}, but this has not yet been extended to the $O(N)$ model, although this extension is technically possible with the graphical function method of Schnetz.

The full 6 loop beta function for an $O(2)$ symmetric theory, as given by \cite{Kompaniets:2017yct}, for the coupling constant $g$ is
\begin{align}\label{eq:BetaO(2)6loops}
    \beta_g &= - \varepsilon g +\frac{10 g^2}{3}-\frac{20 g^3}{3}+g^4 \left(\frac{128 \zeta_3}{9}+\frac{617}{27}\right) \nonumber -g^5 \left(\frac{12160
   \zeta_5}{81}+\frac{8224 \zeta_3}{81}-\frac{640 \zeta_4}{27}+\frac{23927}{243}\right) \nonumber \\
   & +g^6
   \left(\frac{15680 \zeta_7}{9}+\frac{331360 \zeta_5}{243}+\frac{3872 \zeta_3^2}{81}+\frac{168706 \zeta_3}{243}-\frac{5780 \zeta_4}{27}-\frac{38000 \zeta_6}{81}+\frac{103489}{216}\right) \nonumber \\
   &+ g^7 \left(\frac{15488 \zeta_3\zeta_4}{81}-\frac{139479040 \zeta_9}{6561}-\frac{13627136 \zeta_7}{729}-\frac{1495040 \zeta_3 \zeta_5}{243}-\frac{6885760 \zeta_5}{729}\nonumber \right.\\
    &\left. \quad  \quad -\frac{770048 \zeta_3^3}{729}  -\frac{1720832 \zeta_3^2}{729}-\frac{3550148 \zeta_3}{729}-\frac{14336 \zeta_{3,5}}{5}+\frac{399092
   \zeta_4}{243}+\frac{3721600 \zeta_6}{729}\nonumber \right.\\
    &\left.   \quad \quad +\frac{2033024 \zeta_8}{135}-\frac{7345229}{2916}\right)  +\cdots\,.
\end{align}
At 6 loops we now see the appearance of Riemann's  $\zeta_4\approx 1.082323234$, $\zeta_5\approx 1.036927755$, $\zeta_6\approx 1.017343062$, $ \zeta_7\approx 1.008349277$, $\zeta_8\approx 1.004077356$ and $\zeta_9\approx 1.002008393$. As well as the double zeta value $\zeta_{3,5}\approx 0.037707673$. Considering the relationship between the coupling constants for the perfect square and the $O(1,1)$ theory, given by \cref{eq:gPhi4PSequiv}, the beta functions should be equivalent. If we substitute \cref{eq:BetaO(2)6loops} into the equivalence relationship given by \cref{eq:equivalencebetarelationship} we find the beta function for the PS theory to 6 loops is given by
\begin{align}\label{eq:sixloopPS}
\beta_\lambda &= -\frac{\varepsilon}{2}  \lambda-5 \lambda ^3 -30 \lambda ^5- \lambda ^7\left(192 \zeta_3+\frac{617}{2}\right) -\lambda ^9
   \left(6080 \zeta_5 + 4112 \zeta_3-960\zeta_4+\frac{23927}{6}\right) \nonumber\\
   & + \lambda ^{11}\left(-211680 \zeta_7-165680 \zeta_5-5808 \zeta_3^2-84353 \zeta_3+26010\zeta_4 +57000 \zeta_6-\frac{931401}{16}\right)   \nonumber\\
& +\lambda ^{13} \left(69696 \zeta_3\zeta_4-\frac{69739520 \zeta_9}{9}-6813568 \zeta_7 -2242560 \zeta_3 \zeta_5 -3442880 \zeta_5 \right. \nonumber \\
&\left.\quad   \quad -385024 \zeta_3^3 \quad -860416 \zeta_3^2-1775074 \zeta_3-\frac{5225472 \zeta_{3,5}}{5}+598638\zeta_4+1860800 \zeta_6 \right. \nonumber \\
&\left. \quad \quad +\frac{27445824 \zeta_8}{5}-\frac{7345229}{8}\right) +\cdots \,.
\end{align}
If we compare this with our computed beta function result, \cref{eq:BetaThreeloopPS}, we see exact agreement up to three loops. Moreover, as we show in \cite{Anderson:2026Rsqaured}, the one loop results for CFQG agree with the known one loop results for the perfect square and therefore we predict the beta function for CFQG to 6 loops is also given by \cref{eq:sixloopPS}. 

Notably, due to the Ward identity, we are also able to give the anomalous dimension to 6 loops using \cref{eq:sixloopPS} since we find these are related by 
\begin{equation}
    \gamma_\sigma = \frac{\varepsilon}{2} +\frac{\beta_\lambda}{\lambda} \,.
\end{equation}
We note that whilst the running of the couplings are equivalent in the $O(2)$ and PS theories, the anomalous dimensions of the fields are not as easily related.

\section{Conclusion }
\label{sec:Conclusion}

We have reported several advances in the renormalisation of the general, four-dimensional, four-derivative, shift-symmetric scalar field theories identified by Holdom~\cite{Holdom:2023usn}. He showed that these theories are asymptotically free, and yield positive tree-level cross sections in a certain region of coupling constant space~\cite{Holdom:2024cfq}. Here, we have proven several non-renormalisation and IR finiteness theorems. We have also confirmed and extended his calculations of renormalisation constants, field anomalous dimensions, and beta functions up to three loops. 

We began by establishing a structural theorem governing the dependence of the renormalisation counterterms on the cubic and quartic couplings. In particular we noticed that the momentum dependence of the cubic vertex may by written as a Gram determinant, leading to an enhanced external momentum factorisation which was the key to establishing several subsequent results. Therefore, when the quartic coupling is zero, we were able to prove that the cubic interaction is RG invariant, to all orders, and that only the kinetic term is renormalised. One can compare this to the standard massless $\phi^3$-theory, which being super-renormalisable in $D=4$ also has improved UV behaviour, but whose perturbative expansion suffers from IR-divergences already at the off-shell level \cite{Jackiw:1980kv,Larin:2002sc}. We showed that the renormalisation constants obey the non-trivial identity $Z_\sigma Z_3^{3/2}\big|_{\lambda_4 = 0} = 1$, generalising Safari {\it et al.}'s one loop result~\cite{Safari:2021ocb}. It would be interesting to understand whether the observed non-renormalisation of the cubic term can be derived from Galileon symmetry. 

An even more interesting case is the boundary of the physically consistent region identified by Holdom, defined by the curve $\lambda_4=-\frac{1}{2} \lambda_3^2$. Holdom showed at one loop that this trajectory is RG-invariant. On this curve, the classical Lagrangian density takes the form of a perfect square and is reminiscent of Yang-Mills theory in that a single coupling governs the cubic and quartic interactions. We show in a companion paper that this perfect square scalar theory corresponds precisely to the conformally flat limit of quadratic gravity, in which the only non-trivial degree of freedom is the local scale factor of the metric~\cite{Anderson:2026Rsqaured}. The diffeomorphism invariance of quadratic gravity and of its conformally flat limit dictates that the Lagrangian density is the square of a local operator -- the Ricci scalar. In the quantum field theory, it is the Ward identity associated with this local symmetry which protects the perfect square form of the Lagrangian. In Ref.\cite{Anderson:2026Rsqaured} we note that the beta function of the scalar theory agrees with that of the gravitational theory, at one loop. The Ward identity of the gravitational theory then implies a nontrivial all orders relation between the renormalisation constants in the scalar theory, namely $Z_\sigma Z_3^2=  Z_3^2Z_4^{-1} =1$. By explicit calculation in the scalar theory, we have confirmed this relation up to three loops, providing strong evidence for the equivalence of the scalar and gravitational theories. 

The perfect square theory has a positive Euclidean action, making it amenable to non-perturbative (lattice) studies. We anticipate that the beta functions we have calculated will be useful in investigating whether the lattice formulation leads to a well-defined continuum limit. This being so, it will open the way to non-perturbative studies of this special but still highly nontrivial limit of four-dimensional quantum gravity. 

While General relativity is known to be especially well behaved with regards to {\it infrared} divergences, a general concern for higher derivative theories is the potential for these to appear, due to the large, negative powers of momentum in their propagators. However, we were able to prove that all Euclidean correlators in these four-derivative theories are IR finite. This result is non-trivial, although not unexpected. It follows from the derivative structure of the interactions enforced by the shift symmetry. Without this symmetry, IR divergences would appear in dimensional regularisation giving rise to additional $1/\epsilon$ poles. This would signal a breakdown of perturbation theory at large distances even for off-shell correlators. The absence of such divergences therefore further highlights the important role of shift symmetry in yielding a well-defined perturbative description. 

This finding complements our recent demonstration that the perfect square theory, in particular, leads to positive covariant transition probabilities at tree level, in spite of the presence of ghosts. Extending the tree level result to loop level requires handling collinear and soft IR on-shell divergences of the type which are familiar in QCD. The absence of infrared divergences in off-shell amplitudes, in combination with the optical theorem, suggests that it should be possible to resum on-shell IR divergences in the massless theory. For gauge theories, such as QED and QCD the cancellation of IR divergences holds for sufficiently inclusive observables. This is the content of the KLN theorem \cite{Kinoshita:1962ur, PhysRev.133.B1549,Poggio:1976qr, Sterman:1976jh, Frye:2018xjj}. For the special case of processes such as $e^+e^-$ annihilation into hadrons the IR finiteness can in fact be understood more easily \cite{Sterman:1993hfp} as a consequence of the IR finiteness of off-shell correlators and perturbative unitarity. More concretely this is because the cross section is the imaginary part of the photon self energy, which is IR finite off-shell. In the same spirit, our proof of the IR finiteness of off-shell correlators should provide a first important step towards proving the IR finiteness of the entire theory, through a suitable extension of the KLN theorem. The second ingredient required in such a proof, perturbative unitarity, is provided in the forthcoming paper~\cite{Bateman:2026positivity}. 

As mentioned in the introduction, besides their possible relevance to quantum gravity and other puzzles in the Standard Model, higher derivative scalar theories provide a valuable test-bed for calculational methods likely to be useful in both the Standard Model, where it is important to understand effects of higher derivative operators in the operator product expansion, and in effective field theories where higher derivative models are of ubiquitous interest.    
To accomplish the calculations reported here, we have employed two variants of the $R^*$-method described briefly in \cite{deVries:2019nsu} and \cite{Henriksson:2025hwi} as well as a method based purely on an asymptotic momentum expansion \cite{VanThurenhout:2025vui,Falcioni:2024xav,Falcioni:2023luc,Falcioni:2023vqq,Falcioni:2024xyt,Falcioni:2024qpd}. 
Both of these approaches simplify the computation of the renormalisation constants by rendering the potentially very complicated multi-scale multi-loop Feynman integrals into single-scale, massless self-energy type Feynman integrals. The $R^*$ methodology allows one to express the local UV counterterm of $\ell+1$-loop Feynman integrals into those of products of $\ell$-loop integrals. However, this simplification comes at a high cost of additional combinatorial complexity, produced both through extra derivatives and further recursions of counterterms. To tame this complexity we have employed a recently developed momentum-expansion approach, the so-called $R^*_{\textrm{ME}}$ operation. While this greatly improved the performance over the pure $R^*$ approach, it was still too slow to renormalise the 3-loop, 4-point correlator, for which we ended up using the pure asymptotic expansion instead. 

The reason why the four-derivative theory is so challenging to renormalise with $R^*$ is due to both the high powers of momenta in the vertices and in the propagators. While the former require repeated Taylor operators to be employed to lower the degree of divergences of divergent subgraphs, the latter lead to high degree spurious IR divergences, which in turn lead to long recurrences. But even without the overhead of $R^*$, computing the Feynman integrals of the theory is highly challenging, the integration by parts reductions are more complicated and the number of terms per diagram grows fast. In fact we would claim, despite it being a mere scalar theory, that the computations in the theory are more challenging than the respective ones with scalars traded for gluons in Yang Mills Theory, but without the added complications of having to deal with the intricacies of vector fields and spinor and colour algebra. 

Finally, the techniques we have used here may be extended to broader classes of higher-derivative theories and higher derivative observables, including fields other than scalars. This is especially relevant for theories of gravity including higher-order curvature terms. A natural extension would therefore be to investigate whether shift-symmetry can improve IR and UV behaviour for four-derivative tensor fields and whether similar, special RG invariant trajectories appear.

\appendix

\acknowledgments
We thank Kata Benedek, Latham Boyle, John Donoghue, Bob Holdom, Arkady Tseytlin, Vatsalya Vaibhav, Raju Venugopalan and Roman Zwicky for many helpful discussions. MA is supported by the UKRI Centre for Doctoral Training in Algebra, Geometry and Quantum Fields (AGQ), Grant Number EP/Y035232/1. SB is supported by a Higgs Studentship at the School of Physics and Astronomy, University of Edinburgh. FH and NT are supported by the STFC Consolidated Grant ``Particle Physics at the Higgs Centre,'' by the UKRI FLF grant ``Forest Formulas for the LHC'' (Mr/S03479x/1) (FH) and the Higgs Chair at the University of Edinburgh (NT). Perimeter Institute is supported by the Government of Canada, via Innovation, Science and Economic Development, Canada and by the Province of Ontario via the Ministry of Research, Innovation and Science.

\bibliographystyle{jhep}
\bibliography{refs}

\providecommand{\href}[2]{#2}\begingroup\raggedright\begin{thebibliography}{10}

\bibitem{Holdom:2023usn}
B.~Holdom, \emph{{Running couplings and unitarity in a 4-derivative scalar field theory}}, \href{https://doi.org/10.1016/j.physletb.2023.138023}{\emph{Phys. Lett. B} {\bfseries 843} (2023) 138023} [\href{https://arxiv.org/abs/2303.06723}{{\ttfamily 2303.06723}}].

\bibitem{Holdom:2024cfq}
B.~Holdom, \emph{{UV-complete 4-derivative scalar field theory}}, \href{https://doi.org/10.1016/j.nuclphysb.2024.116472}{\emph{Nucl. Phys. B} {\bfseries 1000} (2024) 116472} [\href{https://arxiv.org/abs/2402.09223}{{\ttfamily 2402.09223}}].

\bibitem{Anderson:2026Rsqaured}
M.~Anderson, S.~Bateman, F.~Herzog and N.~Turok, \emph{{The conformally flat limit of Quadratic Gravity}}, {\emph{In preparation} (2026) }.

\bibitem{Coleman:1973freedom}
S.~Coleman and D.~J. Gross, \emph{Price of asymptotic freedom}, \href{https://doi.org/10.1103/PhysRevLett.31.851}{\emph{Phys. Rev. Lett.} {\bfseries 31} (1973) 851}.

\bibitem{Ostrogradsky:1850fid}
M.~Ostrogradsky, \emph{{M{\'e}moires sur les {\'e}quations diff{\'e}rentielles, relatives au probl{\`e}me des isop{\'e}rim{\`e}tres}}, {\emph{Mem. Acad. St. Petersbourg} {\bfseries 6} (1850) 385}.

\bibitem{Woodard:2015zca}
R.~P. Woodard, \emph{{Ostrogradsky's theorem on Hamiltonian instability}}, \href{https://doi.org/10.4249/scholarpedia.32243}{\emph{Scholarpedia} {\bfseries 10} (2015) 32243} [\href{https://arxiv.org/abs/1506.02210}{{\ttfamily 1506.02210}}].

\bibitem{Stelle:1976gc}
K.~S. Stelle, \emph{{Renormalization of Higher Derivative Quantum Gravity}}, \href{https://doi.org/10.1103/PhysRevD.16.953}{\emph{Phys. Rev. D} {\bfseries 16} (1977) 953}.

\bibitem{Fradkin:1981iu}
E.~S. Fradkin and A.~A. Tseytlin, \emph{{Renormalizable asymptotically free quantum theory of gravity}}, \href{https://doi.org/10.1016/0550-3213(82)90444-8}{\emph{Nucl. Phys. B} {\bfseries 201} (1982) 469}.

\bibitem{Avramidi:1985ki}
I.~G. Avramidi and A.~O. Barvinsky, \emph{Asymptotic freedom in higher derivative quantum gravity}, \href{https://doi.org/10.1016/0370-2693(85)90248-5}{\emph{Phys. Lett. B} {\bfseries 159} (1985) 269}.

\bibitem{Riegert:1984kt}
R.~J. Riegert, \emph{{A Nonlocal Action for the Trace Anomaly}}, \href{https://doi.org/10.1016/0370-2693(84)90983-3}{\emph{Phys. Lett. B} {\bfseries 134} (1984) 56}.

\bibitem{Antoniadis:1991fa}
I.~Antoniadis and E.~Mottola, \emph{{4-D quantum gravity in the conformal sector}}, \href{https://doi.org/10.1103/PhysRevD.45.2013}{\emph{Phys. Rev. D} {\bfseries 45} (1992) 2013}.

\bibitem{Boyle:2021jaz}
L.~Boyle and N.~Turok, \emph{{Cancelling the vacuum energy and Weyl anomaly in the standard model with dimension-zero scalar fields}},  \href{https://arxiv.org/abs/2110.06258}{{\ttfamily 2110.06258}}.

\bibitem{Turok:2023amx}
N.~Turok and L.~Boyle, \emph{{A Minimal Explanation of the Primordial Cosmological Perturbations}},  \href{https://arxiv.org/abs/2302.00344}{{\ttfamily 2302.00344}}.

\bibitem{Boyle:2025bxf}
L.~Boyle, N.~Turok and V.~Vaibhav, \emph{{Fixed points of classical gravity coupled with a Standard-Model-like theory}},  \href{https://arxiv.org/abs/2509.09346}{{\ttfamily 2509.09346}}.

\bibitem{Fradkin:1981hx}
E.~S. Fradkin and A.~A. Tseytlin, \emph{{Renormalizable Asymptotically Free Quantum Theory of Gravity}}, \href{https://doi.org/10.1016/0370-2693(81)90702-4}{\emph{Phys. Lett. B} {\bfseries 104} (1981) 377}.

\bibitem{Fradkin:1985am}
E.~S. Fradkin and A.~A. Tseytlin, \emph{Conformal supergravity}, \href{https://doi.org/10.1016/0370-1573(85)90138-3}{\emph{Phys. Rept.} {\bfseries 119} (1985) 233}.

\bibitem{Tseytlin:2017qfd}
A.~A. Tseytlin, \emph{{On divergences in non-minimal $N=4$ conformal supergravity}}, \href{https://doi.org/10.1088/1751-8121/aa920d}{\emph{J. Phys. A} {\bfseries 50} (2017) 48LT01} [\href{https://arxiv.org/abs/1708.08727}{{\ttfamily 1708.08727}}].

\bibitem{Adamo:2018srx}
T.~Adamo, S.~Nakach and A.~A. Tseytlin, \emph{{Scattering of conformal higher spin fields}}, \href{https://doi.org/10.1007/JHEP07(2018)016}{\emph{JHEP} {\bfseries 07} (2018) 016} [\href{https://arxiv.org/abs/1805.00394}{{\ttfamily 1805.00394}}].

\bibitem{Nakach:2018jyu}
S.~Nakach, \emph{{Conformal Higher Spins and Scattering Amplitudes}}, Ph.D. thesis, Imperial Coll., London, 2018.
\newblock 10.25560/66029.

\bibitem{Tseytlin:2022flu}
A.~A. Tseytlin, \emph{{Comments on a 4-derivative scalar theory in 4 dimensions}}, \href{https://doi.org/10.1134/S0040577923120139}{\emph{Theor. Math. Phys.} {\bfseries 217} (2023) 1969} [\href{https://arxiv.org/abs/2212.10599}{{\ttfamily 2212.10599}}].

\bibitem{Bateman:2026letter}
S.~Bateman and N.~Turok, \emph{{Escape from Ostrogradsky via Hidden Ghost Parity}},  \href{https://arxiv.org/abs/2607.00096}{{\ttfamily 2607.00096}}.

\bibitem{Safari:2021ocb}
M.~Safari, A.~Stergiou, G.~P. Vacca and O.~Zanusso, \emph{{Scale and conformal invariance in higher derivative shift symmetric theories}}, \href{https://doi.org/10.1007/JHEP02(2022)034}{\emph{JHEP} {\bfseries 02} (2022) 034} [\href{https://arxiv.org/abs/2112.01084}{{\ttfamily 2112.01084}}].

\bibitem{Symanzik:1973hx}
K.~Symanzik, \emph{{A field theory with computable large-momenta behavior}}, \href{https://doi.org/10.1007/BF02788323}{\emph{Lett. Nuovo Cim.} {\bfseries 6S2} (1973) 77}.

\bibitem{Chetyrkin:1982nn}
K.~G. Chetyrkin and F.~V. Tkachov, \emph{{Infrared R operation and ultraviolet counterterms in the ms scheme }}, \href{https://doi.org/10.1016/0370-2693(82)90358-6}{\emph{Phys. Lett. B} {\bfseries 114} (1982) 340}.

\bibitem{Vladimirov:1979zm}
A.~A. Vladimirov, \emph{{Method for computing renormalization group functions in dimensional renormalization scheme}}, \href{https://doi.org/10.1007/BF01018394}{\emph{Theor. Math. Phys.} {\bfseries 43} (1980) 417}.

\bibitem{Chetyrkin:1984xa}
K.~G. Chetyrkin and V.~A. Smirnov, \emph{{R* operation corrected}}, \href{https://doi.org/10.1016/0370-2693(84)91291-7}{\emph{Phys. Lett. B} {\bfseries 144} (1984) 419}.

\bibitem{Smirnov:1985yck}
V.~A. Smirnov and K.~G. Chetyrkin, \emph{{$R^*$ operation in the Minimal Subtraction Scheme}}, \href{https://doi.org/10.1007/BF01017902}{\emph{Theor. Math. Phys.} {\bfseries 63} (1985) 462}.

\bibitem{Larin:2002sc}
S.~Larin and P.~van Nieuwenhuizen, \emph{{The Infrared R* operation}},  \href{https://arxiv.org/abs/hep-th/0212315}{{\ttfamily hep-th/0212315}}.

\bibitem{Kleinert:2001hn}
H.~Kleinert and V.~Schulte-Frohlinde, \emph{{Critical Properties of $\phi^4$-Theories}}. World Scientific Publishing, 2001, \href{https://doi.org/10.1142/4733}{10.1142/4733}.

\bibitem{Chetyrkin:2017ppe}
K.~G. Chetyrkin, \emph{{Combinatorics of $\mathbf{R}$-, $\mathbf{R^{-1}}$-, and $\mathbf{R^*}$-operations and asymptotic expansions of feynman integrals in the limit of large momenta and masses}},  \href{https://arxiv.org/abs/1701.08627}{{\ttfamily 1701.08627}}.

\bibitem{Herzog:2017bjx}
F.~Herzog and B.~Ruijl, \emph{{The R$^{*}$-operation for Feynman graphs with generic numerators}}, \href{https://doi.org/10.1007/JHEP05(2017)037}{\emph{JHEP} {\bfseries 05} (2017) 037} [\href{https://arxiv.org/abs/1703.03776}{{\ttfamily 1703.03776}}].

\bibitem{deVries:2019nsu}
J.~de~Vries, G.~Falcioni, F.~Herzog and B.~Ruijl, \emph{{Two- and three-loop anomalous dimensions of Weinberg{\textquoteright}s dimension-six CP-odd gluonic operator}}, \href{https://doi.org/10.1103/PhysRevD.102.016010}{\emph{Phys. Rev. D} {\bfseries 102} (2020) 016010} [\href{https://arxiv.org/abs/1907.04923}{{\ttfamily 1907.04923}}].

\bibitem{Beekveldt:2020kzk}
R.~Beekveldt, M.~Borinsky and F.~Herzog, \emph{{The Hopf algebra structure of the R$^{*}$-operation}}, \href{https://doi.org/10.1007/JHEP07(2020)061}{\emph{JHEP} {\bfseries 07} (2020) 061} [\href{https://arxiv.org/abs/2003.04301}{{\ttfamily 2003.04301}}].

\bibitem{Henriksson:2025hwi}
J.~Henriksson, F.~Herzog, S.~R. Kousvos and J.~Roosmale~Nepveu, \emph{{Multi-loop spectra in general scalar EFTs and CFTs}},  \href{https://arxiv.org/abs/2507.12518}{{\ttfamily 2507.12518}}.

\bibitem{Falcioni:2023luc}
G.~Falcioni, F.~Herzog, S.~Moch and A.~Vogt, \emph{{Four-loop splitting functions in QCD {\textendash} The quark-quark case}}, \href{https://doi.org/10.1016/j.physletb.2023.137944}{\emph{Phys. Lett. B} {\bfseries 842} (2023) 137944} [\href{https://arxiv.org/abs/2302.07593}{{\ttfamily 2302.07593}}].

\bibitem{Falcioni:2023vqq}
G.~Falcioni, F.~Herzog, S.~Moch and A.~Vogt, \emph{{Four-loop splitting functions in QCD {\textendash} The gluon-to-quark case}}, \href{https://doi.org/10.1016/j.physletb.2023.138215}{\emph{Phys. Lett. B} {\bfseries 846} (2023) 138215} [\href{https://arxiv.org/abs/2307.04158}{{\ttfamily 2307.04158}}].

\bibitem{Falcioni:2024qpd}
G.~Falcioni, F.~Herzog, S.~Moch, A.~Pelloni and A.~Vogt, \emph{{Four-loop splitting functions in QCD {\textendash} the gluon-gluon case {\textendash}}}, \href{https://doi.org/10.1016/j.physletb.2024.139194}{\emph{Phys. Lett. B} {\bfseries 860} (2025) 139194} [\href{https://arxiv.org/abs/2410.08089}{{\ttfamily 2410.08089}}].

\bibitem{Falcioni:2024xav}
G.~Falcioni, F.~Herzog, S.~Moch and S.~Van~Thurenhout, \emph{{Constraints for twist-two alien operators in QCD}}, \href{https://doi.org/10.1007/JHEP11(2024)080}{\emph{JHEP} {\bfseries 11} (2024) 080} [\href{https://arxiv.org/abs/2409.02870}{{\ttfamily 2409.02870}}].

\bibitem{Falcioni:2024xyt}
G.~Falcioni, F.~Herzog, S.~Moch, A.~Pelloni and A.~Vogt, \emph{{Four-loop splitting functions in QCD {\textendash} The quark-to-gluon case}}, \href{https://doi.org/10.1016/j.physletb.2024.138906}{\emph{Phys. Lett. B} {\bfseries 856} (2024) 138906} [\href{https://arxiv.org/abs/2404.09701}{{\ttfamily 2404.09701}}].

\bibitem{bhabha1950nuclear}
H.~J. Bhabha, \emph{{On a New Theory of Nuclear Forces}}, \href{https://doi.org/10.1103/PhysRev.77.665}{\emph{Phys. Rev.} {\bfseries 77} (1950) 665}.

\bibitem{heisenberg1957lee}
W.~Heisenberg, \emph{{Lee model and quantisation of non linear field equations}}, \href{https://doi.org/https://doi.org/10.1016/0029-5582(87)90060-5}{\emph{Nuclear Physics} {\bfseries 4} (1957) 532}.

\bibitem{Bender:2007wu}
C.~M. Bender and P.~D. Mannheim, \emph{{No-ghost theorem for the fourth-order derivative Pais-Uhlenbeck oscillator model}}, \href{https://doi.org/10.1103/PhysRevLett.100.110402}{\emph{Phys. Rev. Lett.} {\bfseries 100} (2008) 110402} [\href{https://arxiv.org/abs/0706.0207}{{\ttfamily 0706.0207}}].

\bibitem{Bender:2008gh}
C.~M. Bender and P.~D. Mannheim, \emph{{Exactly solvable PT-symmetric Hamiltonian having no Hermitian counterpart}}, \href{https://doi.org/10.1103/PhysRevD.78.025022}{\emph{Phys. Rev. D} {\bfseries 78} (2008) 025022} [\href{https://arxiv.org/abs/0804.4190}{{\ttfamily 0804.4190}}].

\bibitem{Donoghue:2021cza}
J.~F. Donoghue and G.~Menezes, \emph{{On quadratic gravity}}, \href{https://doi.org/10.1393/ncc/i2022-22026-7}{\emph{Nuovo Cim. C} {\bfseries 45} (2022) 26} [\href{https://arxiv.org/abs/2112.01974}{{\ttfamily 2112.01974}}].

\bibitem{Donoghue:2021eto}
J.~F. Donoghue and G.~Menezes, \emph{{Ostrogradsky instability can be overcome by quantum physics}}, \href{https://doi.org/10.1103/PhysRevD.104.045010}{\emph{Phys. Rev. D} {\bfseries 104} (2021) 045010} [\href{https://arxiv.org/abs/2105.00898}{{\ttfamily 2105.00898}}].

\bibitem{Donoghue:2021meq}
J.~F. Donoghue and G.~Menezes, \emph{{Causality and gravity}}, \href{https://doi.org/10.1007/JHEP11(2021)010}{\emph{JHEP} {\bfseries 11} (2021) 010} [\href{https://arxiv.org/abs/2106.05912}{{\ttfamily 2106.05912}}].

\bibitem{Holdom:2015kbf}
B.~Holdom and J.~Ren, \emph{{QCD analogy for quantum gravity}}, \href{https://doi.org/10.1103/PhysRevD.93.124030}{\emph{Phys. Rev. D} {\bfseries 93} (2016) 124030} [\href{https://arxiv.org/abs/1512.05305}{{\ttfamily 1512.05305}}].

\bibitem{Holdom:2021hlo}
B.~Holdom, \emph{{Ultra-Planckian scattering from a QFT for gravity}}, \href{https://doi.org/10.1103/PhysRevD.105.046008}{\emph{Phys. Rev. D} {\bfseries 105} (2022) 046008} [\href{https://arxiv.org/abs/2107.01727}{{\ttfamily 2107.01727}}].

\bibitem{Nicolis:2008in}
A.~Nicolis, R.~Rattazzi and E.~Trincherini, \emph{{The Galileon as a local modification of gravity}}, \href{https://doi.org/10.1103/PhysRevD.79.064036}{\emph{Phys. Rev. D} {\bfseries 79} (2009) 064036} [\href{https://arxiv.org/abs/0811.2197}{{\ttfamily 0811.2197}}].

\bibitem{Kampf:2014rka}
K.~Kampf and J.~Novotny, \emph{{Unification of Galileon Dualities}}, \href{https://doi.org/10.1007/JHEP10(2014)006}{\emph{JHEP} {\bfseries 10} (2014) 006} [\href{https://arxiv.org/abs/1403.6813}{{\ttfamily 1403.6813}}].

\bibitem{Luty:2003vm}
M.~A. Luty, M.~Porrati and R.~Rattazzi, \emph{{Strong interactions and stability in the DGP model}}, \href{https://doi.org/10.1088/1126-6708/2003/09/029}{\emph{JHEP} {\bfseries 09} (2003) 029} [\href{https://arxiv.org/abs/hep-th/0303116}{{\ttfamily hep-th/0303116}}].

\bibitem{Hinterbichler:2015pqa}
K.~Hinterbichler and A.~Joyce, \emph{{Hidden symmetry of the Galileon}}, \href{https://doi.org/10.1103/PhysRevD.92.023503}{\emph{Phys. Rev. D} {\bfseries 92} (2015) 023503} [\href{https://arxiv.org/abs/1501.07600}{{\ttfamily 1501.07600}}].

\bibitem{Goon_2016}
G.~Goon, K.~Hinterbichler, A.~Joyce and M.~Trodden, \emph{Aspects of galileon non-renormalization}, \href{https://doi.org/10.1007/jhep11(2016)100}{\emph{Journal of High Energy Physics} {\bfseries 2016} (2016) }.

\bibitem{Cheung:2014dqa}
C.~Cheung, K.~Kampf, J.~Novotny and J.~Trnka, \emph{{Effective Field Theories from Soft Limits of Scattering Amplitudes}}, \href{https://doi.org/10.1103/PhysRevLett.114.221602}{\emph{Phys. Rev. Lett.} {\bfseries 114} (2015) 221602} [\href{https://arxiv.org/abs/1412.4095}{{\ttfamily 1412.4095}}].

\bibitem{Cheung:2015ota}
C.~Cheung, K.~Kampf, J.~Novotny, C.-H. Shen and J.~Trnka, \emph{{On-Shell Recursion Relations for Effective Field Theories}}, \href{https://doi.org/10.1103/PhysRevLett.116.041601}{\emph{Phys. Rev. Lett.} {\bfseries 116} (2016) 041601} [\href{https://arxiv.org/abs/1509.03309}{{\ttfamily 1509.03309}}].

\bibitem{Brown:2015fyf}
F.~Brown, \emph{{Feynman amplitudes, coaction principle, and cosmic Galois group}}, \href{https://doi.org/10.4310/CNTP.2017.v11.n3.a1}{\emph{Commun. Num. Theor. Phys.} {\bfseries 11} (2017) 453} [\href{https://arxiv.org/abs/1512.06409}{{\ttfamily 1512.06409}}].

\bibitem{maple}
{Maplesoft, a division of Waterloo Maple Inc..}, ``Maple.''

\bibitem{Kuipers:2012rf}
J.~Kuipers, T.~Ueda, J.~A.~M. Vermaseren and J.~Vollinga, \emph{{FORM version 4.0}}, \href{https://doi.org/10.1016/j.cpc.2012.12.028}{\emph{Comput. Phys. Commun.} {\bfseries 184} (2013) 1453} [\href{https://arxiv.org/abs/1203.6543}{{\ttfamily 1203.6543}}].

\bibitem{Vermaseren:2000nd}
J.~A.~M. Vermaseren, \emph{{New features of FORM}},  \href{https://arxiv.org/abs/math-ph/0010025}{{\ttfamily math-ph/0010025}}.

\bibitem{Ruijl:2017dtg}
B.~Ruijl, T.~Ueda and J.~Vermaseren, \emph{{FORM version 4.2}},  \href{https://arxiv.org/abs/1707.06453}{{\ttfamily 1707.06453}}.

\bibitem{Ruijl:2017cxj}
B.~Ruijl, T.~Ueda and J.~A.~M. Vermaseren, \emph{{Forcer, a FORM program for the parametric reduction of four-loop massless propagator diagrams}}, \href{https://doi.org/10.1016/j.cpc.2020.107198}{\emph{Comput. Phys. Commun.} {\bfseries 253} (2020) 107198} [\href{https://arxiv.org/abs/1704.06650}{{\ttfamily 1704.06650}}].

\bibitem{Goode:2024cfy}
J.~Goode, F.~Herzog and S.~Teale, \emph{{OPITeR: A program for tensor reduction of multi-loop Feynman integrals}}, \href{https://doi.org/10.1016/j.cpc.2025.109606}{\emph{Comput. Phys. Commun.} {\bfseries 312} (2025) 109606} [\href{https://arxiv.org/abs/2411.02233}{{\ttfamily 2411.02233}}].

\bibitem{Nogueira:1991ex}
P.~Nogueira, \emph{{Automatic Feynman Graph Generation}}, \href{https://doi.org/10.1006/jcph.1993.1074}{\emph{J. Comput. Phys.} {\bfseries 105} (1993) 279}.

\bibitem{Chakraborty2023}
M.~Chakraborty, \emph{{The Asymptotic Hopf Algebra of Feynman Integrals}},  master's thesis, Indian Institute of Science, 2023.

\bibitem{Chetyrkin:1982zq}
K.~G. Chetyrkin, F.~V. Tkachov and S.~G. Gorishnii, \emph{{Operator product expansion in the minimal subtraction scheme}}, \href{https://doi.org/10.1016/0370-2693(82)90701-8}{\emph{Phys. Lett. B} {\bfseries 119} (1982) 407}.

\bibitem{Chetyrkin:1983qlc}
K.~G. Chetyrkin, \emph{{Infrared R\ensuremath{*} - operation and operator product expansion in the minimal subtraction scheme}}, \href{https://doi.org/10.1016/0370-2693(83)90183-1}{\emph{Phys. Lett. B} {\bfseries 126} (1983) 371}.

\bibitem{Gorishnii:1983su}
S.~G. Gorishnii, S.~A. Larin and F.~V. Tkachov, \emph{{The algorithm for OPE coefficient functions in the MS scheme}}, \href{https://doi.org/10.1016/0370-2693(83)91439-9}{\emph{Phys. Lett. B} {\bfseries 124} (1983) 217}.

\bibitem{Gorishnii:1986gn}
S.~G. Gorishnii and S.~A. Larin, \emph{{Coefficient functions of asymptotic operator expansions in minimal subtraction scheme}}, \href{https://doi.org/10.1016/0550-3213(87)90283-5}{\emph{Nucl. Phys. B} {\bfseries 283} (1987) 452}.

\bibitem{Chetyrkin:1988zz}
K.~G. Chetyrkin, \emph{{Operator expansions in the minimal subtraction scheme. $1$: The gluing method}}, \href{https://doi.org/10.1007/BF01017168}{\emph{Theor. Math. Phys.} {\bfseries 75} (1988) 346}.

\bibitem{Chetyrkin:1988cu}
K.~G. Chetyrkin, \emph{{Operator expansions in the minimal subtraction scheme. $2$: Explicit formulas for coefficient functions}}, \href{https://doi.org/10.1007/BF01028580}{\emph{Theor. Math. Phys.} {\bfseries 76} (1988) 809}.

\bibitem{LlewellynSmith:1987jx}
C.~H. Llewellyn~Smith and J.~P. de~Vries, \emph{{The operator product expansion for minimally subtracted operators}}, \href{https://doi.org/10.1016/0550-3213(88)90407-5}{\emph{Nucl. Phys. B} {\bfseries 296} (1988) 991}.

\bibitem{Gorishnii:1989dd}
S.~G. Gorishnii, \emph{{Construction of operator expansions and effective theories in the MS scheme}}, \href{https://doi.org/10.1016/0550-3213(89)90622-6}{\emph{Nucl. Phys. B} {\bfseries 319} (1989) 633}.

\bibitem{Smirnov:1990rz}
V.~A. Smirnov, \emph{{Asymptotic expansions in limits of large momenta and masses}}, \href{https://doi.org/10.1007/BF02102092}{\emph{Commun. Math. Phys.} {\bfseries 134} (1990) 109}.

\bibitem{Smirnov:1994tg}
V.~A. Smirnov, \emph{{Asymptotic expansions in momenta and masses and calculation of Feynman diagrams}}, \href{https://doi.org/10.1142/S0217732395001617}{\emph{Mod. Phys. Lett. A} {\bfseries 10} (1995) 1485} [\href{https://arxiv.org/abs/hep-th/9412063}{{\ttfamily hep-th/9412063}}].

\bibitem{Smirnov:2002pj}
V.~A. Smirnov, \emph{{Applied asymptotic expansions in momenta and masses}}, \href{https://doi.org/10.1007/3-540-44574-9}{\emph{Springer Tracts Mod. Phys.} {\bfseries 177} (2002) 1}.

\bibitem{Chakraborty:2024uzz}
M.~Chakraborty and F.~Herzog, \emph{{The asymptotic Hopf algebra of Feynman integrals}}, \href{https://doi.org/10.1007/JHEP01(2025)006}{\emph{JHEP} {\bfseries 01} (2025) 006} [\href{https://arxiv.org/abs/2408.14304}{{\ttfamily 2408.14304}}].

\bibitem{VanThurenhout:2025vui}
S.~Van~Thurenhout, G.~Falcioni, F.~Herzog and S.-O. Moch, \emph{{Alien operators for PDF evolution}}, \href{https://doi.org/10.22323/1.485.0465}{\emph{PoS} {\bfseries EPS-HEP2025} (2026) 465} [\href{https://arxiv.org/abs/2509.01994}{{\ttfamily 2509.01994}}].

\bibitem{Boito:2016pwf}
D.~Boito, M.~Jamin and R.~Miravitllas, \emph{{Scheme Variations of the QCD Coupling and Hadronic {\ensuremath{\tau}} Decays}}, \href{https://doi.org/10.1103/PhysRevLett.117.152001}{\emph{Phys. Rev. Lett.} {\bfseries 117} (2016) 152001} [\href{https://arxiv.org/abs/1606.06175}{{\ttfamily 1606.06175}}].

\bibitem{Davies:2017hyl}
J.~Davies and A.~Vogt, \emph{{Absence of $\pi^2$ terms in physical anomalous dimensions in DIS: Verification and resulting predictions}}, \href{https://doi.org/10.1016/j.physletb.2017.11.036}{\emph{Phys. Lett. B} {\bfseries 776} (2018) 189} [\href{https://arxiv.org/abs/1711.05267}{{\ttfamily 1711.05267}}].

\bibitem{Baikov:2018wgs}
P.~A. Baikov and K.~G. Chetyrkin, \emph{{The structure of generic anomalous dimensions and no-$\pi$ theorem for massless propagators}}, \href{https://doi.org/10.1007/JHEP06(2018)141}{\emph{JHEP} {\bfseries 06} (2018) 141} [\href{https://arxiv.org/abs/1804.10088}{{\ttfamily 1804.10088}}].

\bibitem{Abbott:1981ke}
L.~F. Abbott, \emph{{Introduction to the Background Field Method}}, {\emph{Acta Phys. Polon. B} {\bfseries 13} (1982) 33}.

\bibitem{Chetyrkin:1981jq}
K.~G. Chetyrkin, A.~L. Kataev and F.~V. Tkachov, \emph{{Five Loop Calculations in the $g \phi^4$ Model and the Critical Index $\eta$}}, \href{https://doi.org/10.1016/0370-2693(81)90968-0}{\emph{Phys. Lett. B} {\bfseries 99} (1981) 147}.

\bibitem{Gorishnii:1983gp}
S.~G. Gorishnii, S.~A. Larin, F.~V. Tkachov and K.~G. Chetyrkin, \emph{{Five Loop Renormalization Group Calculations in the $g \phi^4$ in Four-dimensions Theory}}, \href{https://doi.org/10.1016/0370-2693(83)90324-6}{\emph{Phys. Lett. B} {\bfseries 132} (1983) 351}.

\bibitem{Kazakov:1983dyk}
D.~I. Kazakov, \emph{The method of uniqueness, a new powerful technique for multiloop calculations}, \href{https://doi.org/10.1016/0370-2693(83)90816-X}{\emph{Phys. Lett. B} {\bfseries 133} (1983) 406}.

\bibitem{Kompaniets:2017yct}
M.~V. Kompaniets and E.~Panzer, \emph{{Minimally subtracted six loop renormalization of $O(n)$-symmetric $\phi^4$ theory and critical exponents}}, \href{https://doi.org/10.1103/PhysRevD.96.036016}{\emph{Phys. Rev. D} {\bfseries 96} (2017) 036016} [\href{https://arxiv.org/abs/1705.06483}{{\ttfamily 1705.06483}}].

\bibitem{Batkovich:2016jus}
D.~V. Batkovich, K.~G. Chetyrkin and M.~V. Kompaniets, \emph{{Six loop analytical calculation of the field anomalous dimension and the critical exponent $\eta$ in $O(n)$-symmetric $\varphi^4$ model}}, \href{https://doi.org/10.1016/j.nuclphysb.2016.03.009}{\emph{Nucl. Phys. B} {\bfseries 906} (2016) 147} [\href{https://arxiv.org/abs/1601.01960}{{\ttfamily 1601.01960}}].

\bibitem{Schnetz:2022nsc}
O.~Schnetz, \emph{{{\ensuremath{\phi}}4 theory at seven loops}}, \href{https://doi.org/10.1103/PhysRevD.107.036002}{\emph{Phys. Rev. D} {\bfseries 107} (2023) 036002} [\href{https://arxiv.org/abs/2212.03663}{{\ttfamily 2212.03663}}].

\bibitem{Schnetz:2016fhy}
O.~Schnetz, \emph{{Numbers and Functions in Quantum Field Theory}}, \href{https://doi.org/10.1103/PhysRevD.97.085018}{\emph{Phys. Rev. D} {\bfseries 97} (2018) 085018} [\href{https://arxiv.org/abs/1606.08598}{{\ttfamily 1606.08598}}].

\bibitem{Jackiw:1980kv}
R.~Jackiw and S.~Templeton, \emph{{How Superrenormalizable Interactions Cure their Infrared Divergences}}, \href{https://doi.org/10.1103/PhysRevD.23.2291}{\emph{Phys. Rev. D} {\bfseries 23} (1981) 2291}.

\bibitem{Kinoshita:1962ur}
T.~Kinoshita, \emph{{Mass singularities of Feynman amplitudes}}, \href{https://doi.org/10.1063/1.1724268}{\emph{J. Math. Phys.} {\bfseries 3} (1962) 650}.

\bibitem{PhysRev.133.B1549}
T.~D. Lee and M.~Nauenberg, \emph{Degenerate systems and mass singularities}, \href{https://doi.org/10.1103/PhysRev.133.B1549}{\emph{Phys. Rev.} {\bfseries 133} (1964) B1549}.

\bibitem{Poggio:1976qr}
E.~C. Poggio and H.~R. Quinn, \emph{{The Infrared Behavior of Zero-Mass Green's Functions and the Absence of Quark Confinement in Perturbation Theory}}, \href{https://doi.org/10.1103/PhysRevD.14.578}{\emph{Phys. Rev. D} {\bfseries 14} (1976) 578}.

\bibitem{Sterman:1976jh}
G.~F. Sterman, \emph{{Kinoshita's Theorem in Yang-Mills Theories}}, \href{https://doi.org/10.1103/PhysRevD.14.2123}{\emph{Phys. Rev. D} {\bfseries 14} (1976) 2123}.

\bibitem{Frye:2018xjj}
C.~Frye, H.~Hannesdottir, N.~Paul, M.~D. Schwartz and K.~Yan, \emph{{Infrared Finiteness and Forward Scattering}}, \href{https://doi.org/10.1103/PhysRevD.99.056015}{\emph{Phys. Rev. D} {\bfseries 99} (2019) 056015} [\href{https://arxiv.org/abs/1810.10022}{{\ttfamily 1810.10022}}].

\bibitem{Sterman:1993hfp}
G.~F. Sterman, \emph{{An Introduction to quantum field theory}}. Cambridge University Press, 8, 1993.

\bibitem{Bateman:2026positivity}
S.~Bateman and N.~Turok, \emph{{Unitarity and positivity in higher-derivative QFTs with hidden ghost parity}}, {\emph{In preparation} (2026) }.

\end{thebibliography}\endgroup

\end{document}